\documentclass[10pt,final,journal,oneside,twocolumn]{IEEEtran}
\usepackage{amssymb}
\usepackage{pstricks}
\usepackage{amsmath}
\usepackage[dvips]{graphicx}
\usepackage{enumerate}
\usepackage{ifthen}
\usepackage{cite}
\newtheorem{theorem}{Theorem}

\newtheorem{lemma}{Lemma}

\IEEEoverridecommandlockouts

\newcounter{MYtempeqncnt}

\begin{document}

\title{Asymptotic Spectral Efficiency of the Uplink in Spatially Distributed Wireless Networks With Multi-Antenna Base Stations}
\author{Siddhartan~Govindasamy,~\IEEEmembership{Member}, D.~W.~Bliss,~\IEEEmembership{Senior Member,~IEEE}, and
  David~H.~Staelin,~\IEEEmembership{Life Fellow,~IEEE \thanks{
S. Govindasamy is with the Franklin W. Olin College of
Engineering. 
(email: siddhartan.govindasamy@olin.edu, d.w.bliss@asu.edu,
staelin@mit.edu). Daniel W. Bliss is at the School of Electrical, Computer and Energy Engineering at
Arizona State University. D. H. Staelin was with the Research Laboratory of
Electronics, Massachusetts Institute of Technology (MIT). Opinions, interpretations, conclusions, and recommendations are those of the
authors and are not necessarily endorsed by the United States Government.
This research was supported in part by the National Science
Foundation under Grant ANI-0333902. Portions of this material
have appeared in Conference Record of the 42nd Asilomar
Conference on Signals, Systems and Computers, October 2008.}} }
\maketitle

\pagenumbering{arabic}

\begin{abstract}
The spectral efficiency of a representative uplink of a given
length, in interference-limited, spatially-distributed wireless
networks with hexagonal cells, simple power control, and
multiantenna linear Minimum-Mean-Square-Error receivers is
found to approach an asymptote as the numbers of base-station
antennas N and wireless nodes go to infinity.  An approximation
for the area-averaged spectral efficiency of a representative
link (averaged over the spatial base-station and mobile
distributions), for Poisson distributed base stations, is also
provided. For large N, in the interference-limited regime, the
area-averaged spectral efficiency  is primarily a function of
the ratio of the product of N and the ratio of base-station to
wireless-node densities, indicating that it is possible to
scale such networks by linearly increasing  the product of the
number of base-station antennas and the relative density of
base stations to wireless nodes, with wireless-node density.
The results are useful for designers of wireless systems with
high inter-cell interference because it provides simple
expressions for spectral efficiency as a function of tangible
system parameters like base-station and wireless-node
densities, and number of antennas. These results were derived
combining infinite random matrix theory and stochastic
geometry.
\end{abstract}


\section{Introduction}

It is increasingly common for multiple wireless networks to be
within interfering distance of each other in urban environments
today  due to proliferation of systems such as city-wide
wireless internet access, pico cells for mobile telephony, and
wireless local-area networks.  Antenna arrays at base stations
that employ spatial interference mitigation can significantly
increase data rates in such systems. It is thus important to
study the spectral efficiencies (b/s/Hz) of wireless links with
multiple antennas in environments that have high base station
or access point and wireless-node densities. In such systems
the densities of nodes (both in-and out-of-cell) and their
distribution in space are important factors as they influence
inter-node distances and hence signal and interference
strengths, which directly impact the
Signal-to-Interference-Plus-Noise-Ratio (SINR), spectral
efficiency and ultimately data rates.

Most works on wireless networks with multi-antenna
base-stations do not explicitly model out-of-cell interference
from spatially distributed in-band interferers which is known
to be very challenging. Andrews et al. \cite{AndrewsCellular}
remark that ``despite decades of research, tractable models
that accurately model other-cell interference (OCI) are still
unavailable, which is fairly remarkable given the size of the
industry".

Several authors have used infinite random matrix theory
techniques similar to ours to analyze multiantenna cellular
networks such as Dai and Poor \cite{PoorMultiCell} and Couillet
et al. \cite{DebbahCorrelatedMIMO}. Neither of these works
models the spatial distribution of nodes and thus do not to
capture the effects of interference from users that are
spatially distributed. Monte-carlo simulations were used in
\cite{HanlyCell} and \cite{Catreux} to analyze small,
spatially-distributed multi-antenna cellular networks. Cellular
networks with \emph{single}-antenna base-stations and spatially
distributed nodes have been analyzed in works such as
\cite{AndrewsCellular}, \cite{HanlyOutageCDMA}, and
\cite{novlan2012analytical}
 using stochastic geometry to model the spatial distribution
of nodes. Further discussion of \cite{AndrewsCellular} and
\cite{novlan2012analytical} which are related to this work are
given at the end of this section. Stochastic geometry has also
been used to study \emph{ad hoc} wireless networks with both
multi and single antenna nodes using both finite and asymptotic
techniques in works such as \cite{AndrewsOutage}, \cite{SPAWC}
\cite{JSACPaper}, \cite{jindal2011multi}, \cite{GagnonMMSE} and
\cite{louie2011open}. Please see \cite{HaenggiJSAC} for a
survey of works utilizing stochastic geometry in both cellular
and ad hoc wireless networks and
\cite{BaccelliStochasticGeometry2} and
\cite{BaccelliStochasticGeometry1} which present an extensive
set of useful stochastic geometry techniques.

In this work, we show that with appropriate normalization, the
spectral efficiency  of a representative uplink in a  network
with hexagonal cells, and base-stations with $N$ antennas using
the linear MMSE receiver converges in probability and derive an
asymptotic expression  for the area-averaged spectral
efficiency. We use the term area-averaged spectral efficiency
to refer to the average spectral efficiency of a link where the
averaging is taken over the locations of all the nodes in the
network and fading, to distinguish it from the ergodic spectral
efficiency in the Shannon sense. Note that the hexagonal-cell
model is an idealized model for base-station placement that is
commonly used in the literature as it offers the best coverage
of the plane if we assume that the coverage associated with
each cell is a disk. Furthermore, with a few modifications, we
apply the techniques developed for hexagonal cells to derive an
approximation to the area-averaged spectral efficiency of a
link in a network where base stations are distributed according
to a Poisson Point Process (PPP).

We consider interference from  spatially distributed in-cell
and out-of-cell wireless nodes that have single antennas and
transmit simultaneously in the same channel using
distance-dependent power control. We assume that signal power
decays with distance according to the standard inverse
power-law model. The area-averaged per-link spectral efficiency
is expressed as a function of the number of receiver antennas
$N$, wireless-node and base-station densities, and path-loss
exponent. While the exact CDF of the spectral efficiency for
finite systems would be ideal, computation of this quantity is
difficult for the uplink in cellular systems with power-control
as the transmit powers of nodes depend on their location on the
plane. Moreover for Poisson distributed base stations, the
transmit powers of mobile users are dependent, further
complicating analysis. We use an asymptotic analysis to handle
complexities of the uplink, in particular the dependence of
transmit powers of the mobile nodes as described in more detail
at the end of this section. The asymptotic techniques also help
handle the difficulties in analytically characterizing the
hexagonal cell model which is typically viewed as being
intractable (as noted in \cite{AndrewsCellular},
\cite{novlan2012analytical}) and are usually analyzed by
Monte-carlo simulation such as in
\cite{XiaoUplinkPowerControl}.

The asymptotic expressions we provide are useful in
understanding the behavior of large networks, such as the rate
of spectral efficiency growth with the number of antennas and
base-station density, and to understand the performance
differences between a network with regularly-spaced, and
completely random base-station placements.

Of the recent works that apply stochastic geometry to analyze
cellular networks, \cite{AndrewsCellular} is of particular note
as they introduce a framework to analyze cellular networks with
Poisson base-station placements. Their work assumes single
antenna nodes, exactly one active wireless node per cell, and
exclusively focuses on the downlink. In their model,
the transmit powers of the base stations are
constant allowing them to use a Poisson shot-noise model for
the interference which is at the heart of the derivation of
their main results. Such an approach is not applicable for the
uplink, which is the focus of this work, due to the correlation
between transmit powers of wireless nodes that result from
power control which is an essential feature of the uplink. The
correlation arises because the transmit powers of the wireless
nodes are dependent on their positions relative to the
base-stations in their respective cells. The size and shape of
the cells are of course dependent. This correlation between
transmit powers precludes applying standard Poisson techniques
which typically require the transmit powers of nodes to be independent
of one another.

As noted in a very recent work by Novlan, et al.
\cite{novlan2012analytical} ``the analysis of the uplink
requires several fundamental changes as compared to the
downlink, nearly all of which make it more challenging." In
\cite{novlan2012analytical} which considers single-antenna
uplinks in random-cell networks, this complexity is handled by
applying certain approximations to the network topology such as
approximating the transmit powers of the wireless nodes as
independent. They make a further approximation on the
base-station distribution by first generating Voronoi cells
about the mobile nodes and then placing a base station
with uniform probability inside each Voronoi cell. Thus, the
base-stations in their model are spatially correlated and not
Poisson. In contrast, in the extension of our results to
Poisson-distributed multi-anntenna base stations, we assume
that cells are formed with the base stations as the generator
points as is typically done (e.g. for the downlink in
\cite{AndrewsCellular}), and that the mobile nodes perform
distance-dependent power control which introduces dependence
between the transmit powers. The associated complexities are
handled by the asymptotic analysis which combines stochastic
geometry and infinite random matrix theory. We validated the
results for finite systems using Monte Carlo simulations that
were also used to characterize the spectral efficiency for a
given outage probability.

\section{System Model} \label{Sec:NetworkModel}

Consider a planar wireless network with base stations
distributed at hexagonal lattice sites with minimum
base-station separation $d$, with a base-station at the origin.
While in practical systems, base-station assignments are based
on strongest received signals rather than distance alone, to
simplify analysis, the wireless nodes  are assumed to
communicate with their closest base station in Euclidian
distance. In other words we assume that the cells are formed by
the Voronoi tessellation of the plane (see e.g. \cite{Stoyan})
with the base stations as the generator points.

The base station at the origin is called the
\emph{representative receiver} which is in a link with a
\emph{representative transmitter} at a distance $r_1$ away. We
shall consider both constant $r_1$ and random $r_1$ resulting
from the representative transmitter being distributed with
uniform probability in the cell associated with the
representative receiver. The later case will be called the
\emph{random link} case.  The link between the two is called
the \emph{representative link}. The representative receiver is
assumed to have $N$ antennas and the representative transmitter
and interferers (to be defined in the next paragraph) have
single antennas.

\begin{figure}
\begin{center}
\includegraphics[width=3.25in]{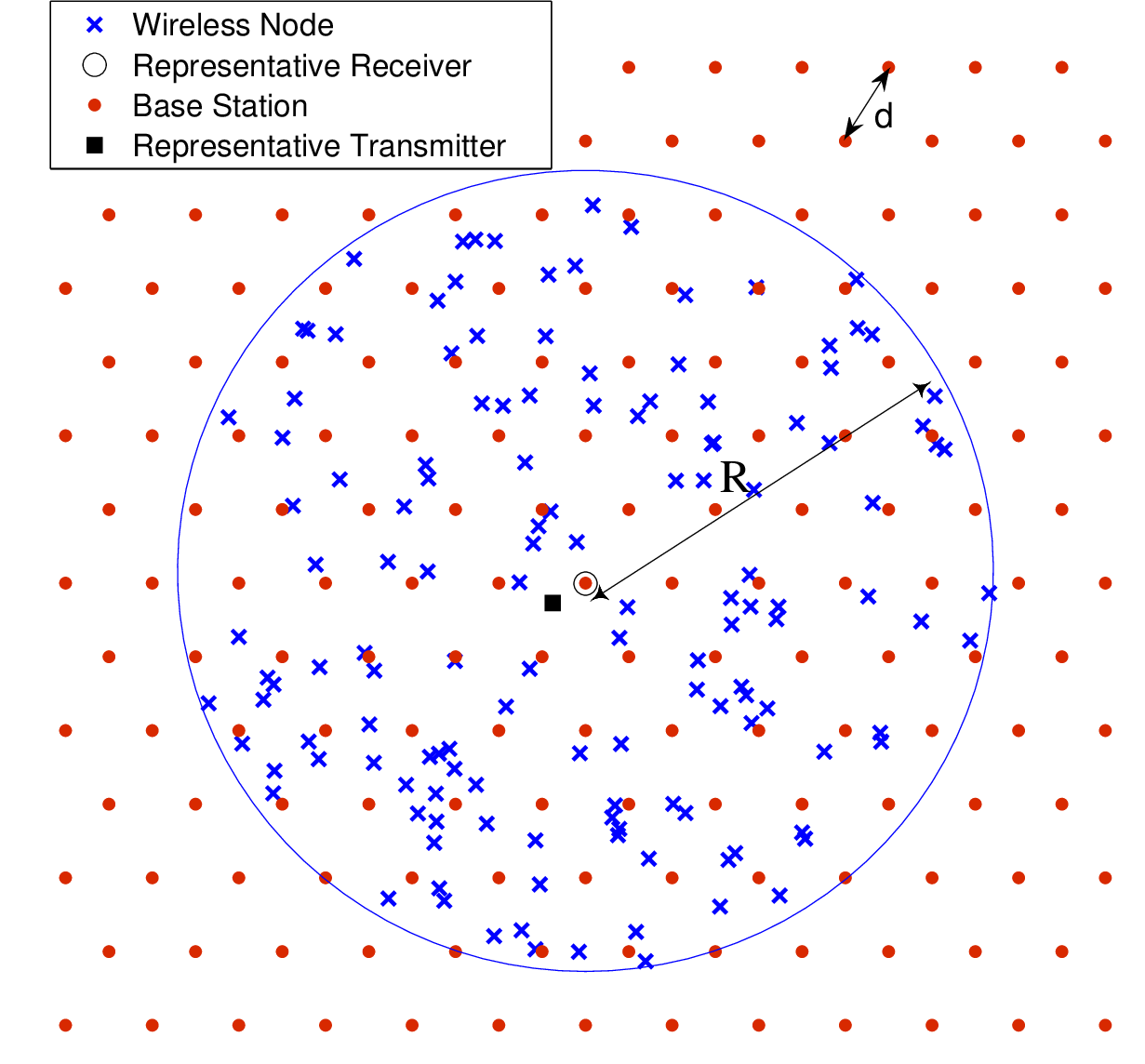}
\caption{Illustration of wireless network with representative link and base-stations at hexagonal lattice sites. The representative receiver is at the origin and the representative transmitter is denoted by the filled square. The remaining base stations are represented by the dots and the interfering wireless nodes are represented with the crosses. \label{Fig:WirelessNetwork} \label{Fig:HexCell}}
\end{center}
\end{figure}

Suppose that there is a circular network of radius $R$ centered
at the origin with $n$ additional wireless nodes (in addition
to the representative transmitter) distributed in an independent, identically distributed
(i.i.d.)
fashion in the network with uniform probability such that
\begin{align}
n = \rho_w \pi R^2 \, ,\label{Eqn:NumberOfNodesRadius}
\end{align}
where $\rho_w$ is the effective area density of the wireless nodes
which are co-channel interferers to the representative link.
Note that these are nodes that are \emph{actively} transmitting
in our model so the true density of nodes may be much higher.
An example of this network is illustrated in Figure
\ref{Fig:HexCell}. The representative transmitter and
interferers are labeled as follows. Node 1 is the
representative transmitter, and nodes $2, 3, \cdots, n+1$ are
the interferers in random order of distance from the origin.
The asymptotic regime we consider is the limit as $N$, $R$ and
$n$ are taken to infinity such that $c = n/N > 2$, $\rho_w$ is
constant and \eqref{Eqn:NumberOfNodesRadius} holds. In the
following we assume that whenever $N\to\infty$, $n$ and $R \to
\infty$ in this manner as well. The resulting network has
wireless nodes distributed uniformly randomly on the entire
plane with density $\rho_w$. Note that since we are interested
in large wireless networks with moderately large numbers of
base-station antennas, $2\ll c$.

The $i$-th wireless node is at distance $r_i$ from the
representative receiver at the origin and is assumed to
transmit with power $P_i$. The average received signal power
per antenna (averaged over the fading distribution defined in
the next paragraph) due to the $i$-th wireless node is
\begin{align}
p_i = P_i \, G_t\,r_i ^{-\alpha}
\end{align}
with the path-loss exponent $\alpha > 2$, and $G_t$ is a
proportionality constant. The wireless nodes control their
transmit power in order to achieve a target received power
relative to path loss at their closest base station, subject to
a maximum power constraint, $P_M$ as follows
\begin{align}
P_i =  \min\left(\frac{p_t}{G_t}r_{ti}^\alpha, P_M\right)\,, \label{Eqn:PowerControlHex}
\end{align}
where $r_{ti}$ is the distance between the $i$-th wireless node
and its closest base station. Let the limiting probability density function (PDF) of $P_i$ be denoted
by $f_P(p)$ and $E[P^{\frac{2}{\alpha}}]$ be its expected value
raised to the power $\frac{2}{\alpha}$.

We assume frequency-flat fading with independent, circularly
symmetric complex Gaussian channel coefficients between all
pairs of antennas.  Let $\mathbf{y} \in \mathbb{C}^{N\times 1}$
be the vector of sampled received signals at the $N$ antennas of
the representative receiver at a given sampling time, and
$\mathbf{w}\in \mathbb{C}^{N\times 1}$ contain zero-mean,
i.i.d. complex Gaussian noise terms of variance ${\sigma}^2$
denoted by $\mathcal{CN}(0,{\sigma}^2)$ .  This system can be
represented by the following  equation:
\begin{align}
\label{Eqn:BasicLinearEquationTethered}
\mathbf{y} = \sqrt{p_1}\,\mathbf{g}_{1}\, x_1 + \sum_{i = 2}^{n+1} \sqrt{p_i}\, \mathbf{g}_{i}\, x_i + \mathbf{w}
\end{align}
where $\mathbf{g}_i\in\mathbb{C}^{N \times 1}$ has i.i.d.
$\mathcal{CN}(0,1)$ entries and $x_i$ is the transmitted symbol
of the $i$-th wireless node with $E[|x_i|^2] =1$. Thus,
$\mathbf{g}_i$ captures the Rayleigh fading and $p_i$ captures
the  combined transmit power and path loss associated with
node-$i$. To focus on the interference-limited regime, we set
the noise power $\sigma^2=0$.

We assume that the base stations use spatial linear MMSE
estimators to mitigate interference. Note that the linear MMSE
receiver is the linear receiver  that maximizes the SINR (e.g.
see \cite{TseHanly}) which maximizes the spectral efficiency
for Gaussian signals. We assume that all nodes  use Gaussian
codebooks which results in Gaussian distributed residual
interference at the output of the linear MMSE receiver. Thus,
the spectral efficiency is given by the Shannon formula as is
commonly done in the literature (e.g. \cite{AndrewsCellular}).
It is important to note here that the rapid decay of signal
power with distance associated with the inverse power-law
path-loss model means that the central-limit theorem does not
hold for a general distribution of transmit signals (e.g.
Quadrature-Amplitude-Modulation) \cite{Gulati}. Thus the
aggregate interference at the \emph{input} to the MMSE receiver
will not be Gaussian distributed  (e.g. see \cite{Gulati}) if
the transmitted signals themselves are not Gaussian to begin
with. If we do not make the assumption that the transmitted
signals are Gaussian distributed, the spectral efficiencies we
compute should be interpreted as \emph{achievable} spectral
efficiencies because the Gaussian distribution is entropy
maximizing. Thus from an information theoretic perspective, the
spectral efficiency obtained by assuming a Gaussian
interference distribution is a lower bound to the spectral
efficiency achievable with any other interference distribution.
Additionally, it is common practice to design systems to
operate in Gaussian noise. One could apply a
correction factor, $\eta$ say, to the SIR and compute the spectral
efficiency as $log_2(1+\eta \mbox{SIR})$. This has been
suggested in \cite{AndrewsCellular} and other works. While we
do not use a scale factor of $\eta$ here, introducing it into
our expressions is straightforward.

The main results of this work will be given in terms of a
normalized version of the Signal-to-Interference-Ratio (SIR),
\begin{align}
\beta_N &= N^{-{\frac{\alpha}{2}}} \, \mathbf{g}_1^\dagger \left(\sum_{i = 2}^{n+1} p_i\,\mathbf{g}_i\,\mathbf{g}_i^\dagger \right)^{-1}  \mathbf{g}_1\, \;\;\mbox{for which} \label{Eqn:SIROrigDef}\\
\mbox{SIR} &= p_1\, N^{{\frac{\alpha}{2}}}\,\beta_N\,.
\end{align}
Note that up to the normalization by $N^{-\alpha/2}$,
\eqref{Eqn:SIROrigDef} is the standard equation for the SINR
associated with a linear MMSE receiver with the noise variance
assumed to equal zero, as we have assumed here. This assumption
is used in order to utilize an asymptotic approach to
characterize interference-limited systems. We make the
additional observation here that although we assume zero noise,
the resulting receiver does not reduce to a zero-forcing
receiver as the number of antennas $N$ is less than the number
of interferers since $c = n/N > 2$ by assumption. This means
that the degrees of freedom at the receiver are insufficient to
force the interference to zero.

Note that the normalization of the SIR by
$N^{{\frac{\alpha}{2}}}$ keeps the SIR finite as $N\to\infty$
because the SIR grows as $N^{\frac{\alpha}{2}}$. This order of
growth of the SIR with the number of antennas in networks with
the inverse-power-law path-loss model is known and can be
interpreted intuitively  as is done for ad hoc networks in
\cite{JSACPaper}, or using a precise analysis as done in
\cite{jindal2011multi}. Based on our description in
\cite{JSACPaper}, note that the representative receiver can use
a fraction of its degrees of freedom to null nearby interferers
who occupy a disk of radius on the order of $\sqrt{N}$ around
the representative receiver. The aggregate interference from
the un-nulled interferers outside this disk is of order
$N^{\alpha/2-1}$. The remaining fraction of the degrees of
freedom are used to add signals from the target transmitter
coherently, increasing signal power relative to interference by
a factor on the order of $N$. The combined effect is that the
SIR grows as a factor of $N^{\alpha/2}$.

\section{Main Results}
The main results of this work are based on the following
theorem proved in Appendix \ref{Sec:MainTheoremProof} using
Lemma \ref{Lemma:SIRConvergenceLemma} which follows.
\begin{theorem} \label{Theorem:TetheredSIR}
Consider the network model from Section \ref{Sec:NetworkModel}.
As  $N, n, R \to \infty$, the normalized SIR, $\beta_N$
converges in probability to a limit $\beta$ which is the unique
non-negative solution to the following equation
\begin{align}
&E[P^{{\frac{2}{\alpha}}}]\beta^{{\frac{2}{\alpha}}} \left[\frac{\pi}{\alpha} \csc \left({\frac{2\pi}{\alpha}}\right)\right] - \frac{2\pi\rho_w \beta}{\alpha }\times \nonumber \\
&\;\;\;\;\;\;\;\;\;\;\;\;\;\;\int_{0}^{\infty}\frac{\tau^{-\frac{2}{\alpha}}}{1+\tau \beta}  \int_{\tau/b}^\infty f_P(x)x^{\frac{2}{\alpha}} dx\, d\tau = \frac{1}{2\rho_w\pi}  \label{Eqn:TheoremTetheredSINR}
\end{align}
\vspace{0.25cm} where
$b=\left(\frac{\pi\rho_w}{c}\right)^{\frac{\alpha}{2}}$.
\end{theorem}
\begin{lemma} \label{Lemma:SIRConvergenceLemma}
Consider the quantity
\begin{align}
\gamma_N = \frac1N \mathbf{s}^\dagger\left(\frac1N\mathbf{S \Psi S}^\dagger\right)^{-1}\mathbf{s} \label{Eqn:NormSIRDefGenLemma}
\end{align}
where $\mathbf{s} \in \mathbb{C}^{N\times 1}$ and $\mathbf{S}
\in \mathbb{C}^{N\times n}$ comprise i.i.d., zero-mean,
unit-variance entries from a continuous distribution, $n/N = c$, and  $\mathbf{\Psi} =
\mbox{diag}(\psi_2, \psi_3, \cdots \psi_{n+1})$. Note that
$\mathbf{R} = \frac1N\mathbf{S \Psi S}^\dagger$ is invertible
with probability 1 ($w.p.1$) since $N < n$. Suppose that as $n,
N \to\infty$, the empirical distribution function (e.d.f.) of
the diagonal entries of $\mathbf{\Psi}$ converges $w. p. 1$ to
a function $H(x)$. Additionally, assume that there exists an
$N_0$ such that $\forall n
> N_0$ , the minimum eigenvalue of
$\mathbf{R}$ is bounded from below by $\lambda_{\ell b} > 0$,
$w. p. 1$. Then, $\gamma_N \to \gamma$ in probability where
$\gamma$ is given by the non-negative real solution for $m$ in
\begin{equation} \label{Eqn:FixedPoint} 1 =
m\,c\int_{0}^{\infty} \frac{\tau dH(\tau)}{1+\tau m}\,.
\end{equation}
\end{lemma}
\noindent{\it Proof:} Please see Appendix \ref{Lemma:SIRConvergenceLemmaProof}.

Note that Lemma \ref{Lemma:SIRConvergenceLemma} is closely
related to several results in the literature concerning the
convergence of the SINR  of random Direct/Sequence
Code-Division-Multiple-Access (DS/CDMA) systems such as
\cite{TseHanly} and \cite{BaiSilversteinMIMOCDMA}. The existing
results however assume that the noise power is strictly
positive and are thus not directly applicable to systems with
negligible noise power. Lemma \ref{Lemma:SIRConvergenceLemma}
is proved by modifying the proof in \cite{TseHanly}, replacing
the requirement of the strictly positive noise by the
requirement on the minimum eigenvalue of the matrix
$\mathbf{R}$.

Since we assume that all nodes use Gaussian codebooks which
results in Gaussian residual interference at the output of the
MMSE receiver, we can approximate the per-link spectral
efficiency using the Shannon formula assuming that the noise is
negligible as follows.
\begin{align*}
C(r_1) = \log_2(1+\mbox{SIR})= \log_2(1+N^{{\frac{\alpha}{2}}}\, P_1\,r_1^{-\alpha}\,\beta_N)\,,
\end{align*}
where we emphasize that the spectral efficiency is a function
of the length of the representative link, $r_1$. Note that if
the transmit signals are not Gaussian, as we noted in the
system model, the spectral efficiency above and in subsequent
expressions should be interpreted as achievable spectral
efficiencies.

Since the log function is continuous, as $N \to\infty$
$\beta_N\to\beta$ and,
\begin{align}
C(r_1) - \log_2(N^{{\frac{\alpha}{2}}}) \to \log_2( P_1\,r_1^{-\alpha}\,\beta)\,,
\end{align}
in probability (e.g. see \cite{Karr}). Hence, with appropriate
normalization, the spectral efficiency approaches  an asymptote
as $N \to \infty$.  We define this asymptotic spectral
efficiency as  $C^*(r_1) = \log_2(1+N^{{\frac{\alpha}{2}}}
P_1\,r_1^{-\alpha}\,\beta)$.

While $\beta$ is given implicitly by Theorem 1\footnote{To the
best of our knowledge, the only scenario in which similar
techniques have resulted in closed form solutions are when the
received interference powers from all users are equal
\cite{TseHanly}} and has to be solved numerically, we can
approximate the spectral efficiency of a system where the
number of interferers $n$ greatly exceeds the number of
base-station antennas $N$, i.e. small $b$ because the second
term on the LHS of \eqref{Eqn:TheoremTetheredSINR} is small in
this case. In fact this term can be shown to go to zero as
$b\to 0$ (after the limits on $n,N$ and $R$ are taken)
\cite{JSACPaper}. Writing $G_\alpha = \left[\frac{\alpha}{2\pi}
\sin\left({\frac{2\pi}{\alpha}}\right)\right]^{\frac{\alpha}{2}}$,
this yields the following approximation
\begin{align}
 C^*(r_1) \approx \log_2\left(1+N^{{\frac{\alpha}{2}}} P_1\,G_\alpha\left(\frac{1}{E[P^{{\frac{2}{\alpha}}}]\pi\rho_w r_1^2}\right)^{{\frac{\alpha}{2}}}\right)\,. \label{Eqn:AsymptoticSpecEff}
\end{align}
Applying the dominated convergence theorem with steps similar
to that in Appendix E of \cite{LimitedTxCSI} with the noise
power replaced by $\lambda_{\ell\,b}$ we can show that
\begin{align}
E[C|r_1] - C^*(r_1) \to 0\,. \label{Eqn:MeanSpecEffDev}
\end{align}
Hence the asymptotic spectral efficiency $C^*$ is a good
approximation for the conditional area-averaged spectral
efficiency $E[C|r_1]$ (averaging is over wireless-node
locations and fading distributions here) for large $N$.
Finally, we can find the unconditioned area-averaged spectral
efficiency of a random link by averaging with respect to the
distribution of $r_1$ so that
\begin{align}
E[C] \approx \int C(r_1) f_{r_1}(r)\,dr \,, \label{Eqn:GeneralMeanSpecEff}
\end{align}
where $f_{r_1}(r)$ is the PDF of $r_1$ which equals $f_X(x)$
given in Lemma \ref{Lemma:HexLinkLengthLemma}.

If the minimum distance between base stations $d \leq
\frac{3}{\sqrt{3}}
\left(\frac{G_tP_{M}}{p_t}\right)^{\frac{1}{\alpha}}$, the
cells are small enough that all wireless nodes have sufficient
transmit power to meet the target received power $p_t$ at their
base-stations. We call this the sufficient-power case in which
the asymptotic spectral efficiency takes a simple form which is
independent of whether $r_1$ is fixed or random due to the
power control. Substituting the power control equation
\eqref{Eqn:PowerControlHex} into \eqref{Eqn:AsymptoticSpecEff}
\begin{align}
&E[C] \approx  C^*(r_1) \nonumber \\
&\approx \log_2\left(1+\frac{p_t}{G_t}r_{1}^\alpha G_\alpha \left(\frac{N}{E\left[\left(\frac{p_t}{G_t}r_{ti}^\alpha\right)^{\frac{2}{\alpha}}\right]\pi\rho_w r_1^2}\right)^{\frac{\alpha}{2}}\right)
\end{align}
\begin{align}
 &= \log_2\left(1+ G_\alpha
\left(\frac{N}{E\left[r_{ti}^2\right]\pi\rho_w
}\right)^{\frac{\alpha}{2}}\right)\,,
\label{Eqn:TetheredNodeMeanSpecEffNoPowLimit}
 \end{align}
which is a  function of the second moment of the distance
between a random wireless node and its closest base station
$E[r_{ti}^2]$, wireless-node density $\rho_w$, number of
antennas $N$ and path-loss exponent. $E[r_{ti}^2]$ can be found
using the following lemma that statistically characterizes the
distance between a random wireless node and its closest base
station.

\begin{lemma} \label{Lemma:HexLinkLengthLemma}
The PDF $f_X(x)$, CDF $F_X(x)$, and $k$-th moment of the link
length $x$ between a randomly located wireless node and its
closest base station in a hexagonal-cellular system with
minimum base-station separation  $d$ are the following:
\begin{align}
f_X(x) &=
\begin{cases}
\frac{4\pi}{\sqrt{3}d^2}x, &\text{if $ 0 <  x  < \frac{d}{2}$}
\\
\frac{4\pi}{\sqrt{3}d^2}x - \frac{8\sqrt{3} x}{d^2}\cos^{-1}\left(\frac{d}{2x}\right), & \text{if $ \frac{d}{2} <  x  < \frac{\sqrt{3}d}{3}$}
\\
 0, & \text{otherwise.}
\end{cases}\label{Eqn:HexDistanceDensity}\\
%
F_X(x) &=
\begin{cases} 0, & \text{if $x<0$,}
\\
\frac{2\sqrt{3}\pi x^2}{3d^2}, &\text{if $ 0 \le  x  < \frac{d}{2}$}
\\
\frac{2\sqrt{3}\pi x^2}{3d^2} - \frac{4\sqrt{3} x^2}{d^2}\cos^{-1}\left(\frac{d}{2x}\right) & \\
\;\;\;\;\;\;\;\;\;\;\;\;\;\;\;+ 2\sqrt{3}\left(\frac{x^2}{d^2} - \frac14\right)^{\frac12}, & \text{if $ \frac{d}{2} \le  x  < \frac{\sqrt{3}d}{3}$}
\\
1, &\text{if $x \geq
\frac{\sqrt{3}d}{3}$.}\label{Eqn:HexDistanceDist}
\end{cases}\\
E(x^k) &= \frac{2\sqrt{3}}{k+2}\left(\frac{d}{2}\right)^k\int_0^{\frac\pi 6} \frac{1}{\left(\cos \tau\right)^{k+2}}\,d\tau \,.\label{Eqn:HexMoments}\;\;\;\;\;\;
\end{align}

\begin{figure}
\begin{center}
\includegraphics[width=2.5in]{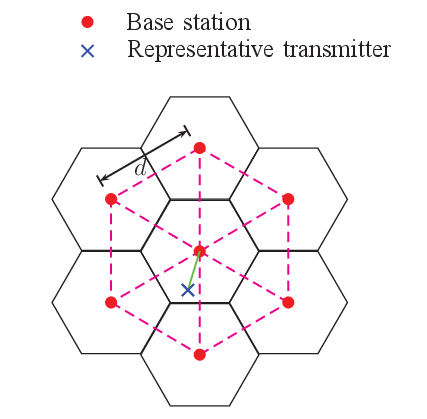}
\caption{Illustration of base stations at hexagonal lattice sites. \label{Fig:HexGrid}}
\end{center}
\end{figure}

\begin{proof}
Consider Figure \ref{Fig:HexGrid} which illustrates a portion
of a wireless network with hexagonal cells. Each wireless node
in the network falls on some random point in an equilateral
triangle formed by the three base stations closest to it, and
forms a link with the base station at the closest vertex of
that triangle as illustrated in Figure \ref{Fig:HexGrid}. Thus,
the link-lengths are statistically equivalent to the distance
between a randomly selected point in an equilateral triangle to
the closest vertex of that triangle. The CDF, PDF and k-th
moments  of the distance between a random point in an
equilateral triangle to the closest vertex  are known
\cite{Mathai}, and are  precisely the formulae in Lemma
\ref{Lemma:HexLinkLengthLemma}. Note that the PDF of
link-lengths associated with a hexagonal cell which
equals  \eqref{Eqn:HexDistanceDensity}, has been given without
proof before in \cite{HexDistanceRef}.
\end{proof}
\end{lemma}

From \eqref{Eqn:TetheredNodeMeanSpecEffNoPowLimit}, the
spectral efficiency  depends on the second moment of
link-lengths  given by \eqref{Eqn:HexMoments} with $k = 2$.
\begin{align*}
E[t_{ti}^2] = \frac{5}{36}d^2 \approx 0.14 d^2.
\end{align*}
Substituting into \eqref{Eqn:TetheredNodeMeanSpecEffNoPowLimit}
yields the following  approximation for the area-averaged and
asymptotic uplink spectral efficiency of a random link (as
defined in Section \ref{Sec:NetworkModel}), in
interference-limited, hexagonal-cell systems with a large
number of base-station antennas and high transmit power
budgets:
\begin{align}
 E[C] \approx C^*(r_1) \approx \log_2\left(1+ G_\alpha \left(\frac{36\, N}{5 \, d^2 \pi\rho_w }\right)^{\frac{\alpha}{2}}\right).\label{Eqn:HexSufficientPower}
\end{align}
In terms of the effective density of base stations from the
hexagonal-cell model $\rho_h$ we can write,
\begin{align}
E[C] \approx C^*(r_1) \approx \log_2\left(1+ G_\alpha \left(\frac{1.98\, N \rho_h}{ \rho_w }\right)^{\frac{\alpha}{2}}\right).\label{Eqn:HexSufficientPowerByDensity}
\end{align}

If $d
> \frac{3}{\sqrt{3}}\left(\frac{G_tP_{M}}{p_t}\right)^{\frac{1}{\alpha}}$,
the transmit power budget is insufficient for all nodes to meet
the target received power at their base stations  which results
in some  wireless nodes transmitting at full power.  In this
case, $E[P^{{\frac{2}{\alpha}}}]$ is given by the following
lemma which can be proved by direct computation using Lemma
\ref{Lemma:HexLinkLengthLemma}:
\begin{lemma} \label{Eqn:LemmaInterfContributionTetheredHex}
If $ P_M < \frac{p_t}{G_t}\left(\frac{d}{2}\right)^\alpha $,
\begin{align}
E[P^{\frac{2}{\alpha}}] =   P_M^{\frac{2}{\alpha}} -  \frac{\sqrt{3}\pi}{3d^2} \left(\frac{G_t}{p_t}\right)^{\frac{2}{\alpha}}P_M^{\frac{4}{\alpha}}. \label{Eqn:TetheredNodeHighlyInsufficientPower}
\end{align}
 If $\frac{p_t}{G_t}\left(\frac{d}{2}\right)^\alpha \leq P_M < \frac{p_t}{G_t}\left(\frac{\sqrt{3}}{3}d\right)^\alpha$,
\begin{align}
&E[P^{\frac{2}{\alpha}}] =    P_M^{\frac{2}{\alpha}} -  \frac{\pi\sqrt{3}}{3d^2}\left(\frac{p_t}{G_t}\right)^{-\frac{2}{\alpha}}P_M^{\frac{4}{\alpha}}  \nonumber \\
&\;\;+\frac{2\sqrt{3}}{d^2}\left(\frac{p_t}{G_t}\right)^{-\frac{2}{\alpha}}P_M^{\frac{4}{\alpha}}\cos^{-1} \left( {\frac {d}{2}\left(\frac{p_t}{G_tP_M}\right)^{\frac{1}{\alpha}}}\right) \nonumber \\
&\;\;+\left(\frac{\sqrt{3}d}{12} \left(\frac{p_t}{G_t}\right)^{\frac{2}{\alpha}}-\frac{5\sqrt{3}}{6d}P_M^{\frac{2}{\alpha}}\right) \sqrt{4\left(\frac{G_tP_M}{p_t}\right)^{\frac{2}{\alpha}}-d^2}. \label{Eqn:TetheredNodeModerateInsufficientPower}
\end{align}
\end{lemma}
Lemma \ref{Eqn:LemmaInterfContributionTetheredHex} substituted
into \eqref{Eqn:AsymptoticSpecEff} yields the area-averaged
spectral efficiency for a length $r_1$ link.

\begin{figure*}[!t]

\normalsize
\setcounter{MYtempeqncnt}{\value{equation}}
\setcounter{equation}{23}
\begin{align}
&E[C] \approx  \int_0^{r_t} \log_2\left(1+G_\alpha \frac{p_t}{G_t}x^{\alpha} \left(\frac{N}{E[P^{\frac{2}{\alpha}}]\pi\rho_wx^2}\right)^{\frac{\alpha}{2}}\right) f_X(x)dx + \int_{r_t}^{\frac{\sqrt{3}d}{3}} \log_2\left(1+G_{\alpha}P_M \left(\frac{N}{E[P^{\frac{2}{\alpha}}]\pi\rho_wx^2}\right)^{\frac{\alpha}{2}}\right) f_X(x) dx \nonumber
\\
&=
F_X\left(r_t\right)\log_2\left(1+
G_\alpha\frac{p_t}{G_t}\left(\frac{N}{E[P^{\frac{2}{\alpha}}]\pi\rho_w}\right)^{\frac{\alpha}{2}}\right)
+\int_{r_t}^{\frac{\sqrt{3}d}{3}}
\log_2\left(1+G_\alpha P_M
\left(\frac{N}{E[P^{\frac{2}{\alpha}}]\pi\rho_wx^2}\right)^{\frac{\alpha}{2}}\right)
f_X(x) dx \label{Eqn:HexNodesInsuffPowerMean}
\end{align}
\vspace*{4pt}
\hrulefill
\setcounter{equation}{\value{MYtempeqncnt}}
\addtocounter{equation}{2}
\end{figure*}

Averaged over the PDF of link-lengths arising from hexagonal
cells (i.e., the representative transmitter is distributed with
uniform probability in the cell containing the origin), the
area-averaged spectral efficiency of a random link is given by \eqref{Eqn:HexNodesInsuffPowerMean} at the top of this page,
where we have used $r_t =\left(\frac{p_t}{P_MG_t}\right)^{-\frac{1}{\alpha}}$,
$F_X(x)$ and $f_X(x)$  from Lemma 1, and
$E[P^{{\frac{2}{\alpha}}}]$  is from Lemma
\ref{Eqn:LemmaInterfContributionTetheredHex}. We were not able
to integrate the second term on the RHS of
\eqref{Eqn:HexNodesInsuffPowerMean} in closed form and thus use numerical
integration to compute it.

\section{Extension to Poisson Distributed Base Stations}

\subsection{Area-averaged Spectral Efficiency}
 The results for the hexagonal-cell model can be
extended to a
 \emph{Poisson-cell} model where base stations are distributed
according to a homogenous PPP with density $\rho_t$,
conditioned on there being a point of the PPP at the origin. We
denote the conditioned  PPP by PPP$^o$. Conditioned on a realization of the
base-station locations $\Pi_t$, Theorem 1 still holds if
$E[P^{\frac{2}{\alpha}}]$ and $f_P(p)$ are replaced with
$E[P^{\frac{2}{\alpha}}|\Pi_t]$ and $f_{P|\Pi_t}(p|\Pi_t)$
respectively.

The ergodicity of the PPP however implies that
$E[P^{\frac{2}{\alpha}}|\Pi_t]$ is equal for almost all realizations of
$\Pi_t$ (i.e. any deviations occur with probability zero).
Similarly, $f_{P|\Pi_t}(p|\Pi_t)$ is equal for almost all
realizations of $\Pi_t$. These properties and the expressions
for  $E[P^{\frac{2}{\alpha}}|\Pi_t]$  and $f_{P|\Pi_t}(p|\Pi_t)$  are
given explicitly in the following lemma.
\begin{lemma} \label{Lemma:ErgodicSpatialAverage}
With probability 1,
\begin{align}
E[P^{\frac{2}{\alpha}}|\Pi_t] &= \!\left( \frac{p_t}{G_t}\right)^{\frac{2}{\alpha}}\frac{1-e^{-\pi\rho_t{\left(\frac{G_t}{p_t}P_M\right)}^{\frac{2}{\alpha}}}}{\pi\rho_t} \nonumber \\
&= E[P^{\frac{2}{\alpha}}] = E[P^{\frac{2}{\alpha}}|\Pi_t,r_1]. \label{Eqn:TetheredNodePowerScale}
\end{align}
and
\begin{align}
&f_{P|\Pi_t}(p|\Pi_t)=f_{P|\Pi_t,r_1}(p|\Pi_t, r_1)=\nonumber \\
& f_P(p)=\begin{cases}
\frac{2 \rho_t\,\pi}{\alpha p}
\left(\frac{p\,G_t}{p_t}\right)^{\frac{2}{\alpha}}e^{-\rho_t \,\pi
\left(\frac{p\,G_t}{p_t}\right)^{\frac{2}{\alpha}}}
, &\text{if  $p \leq P_M$}
\\
 0, & \text{otherwise.}
 \end{cases}\label{Eqn:ErgodicPowerDens}
\end{align}
\end{lemma}
{\it Proof:} In Appendix
\ref{Sec:ErgodicSpatialAverageLemmaProof}.

Thus, the solution for $\beta$ in Theorem 1 takes a fixed value
for almost all realizations of $\Pi_t$. From
\eqref{Eqn:MeanSpecEffDev} the area-averaged spectral
efficiency conditioned on $\Pi_t$ and $r_1$, $E[C|\Pi_t, r_1]$,
has the following property $w.p.1$ as $n,N,R\to\infty$.
\begin{align}
E[C|\Pi_t, r_1] - \log_2(1+ P_1 r_1^{-\alpha}N^{\alpha/2}\beta) \to 0, \label{Eqn:CondMSEPoisson}
\end{align}
where $\beta$ is the non-negative solution to
\eqref{Eqn:TheoremTetheredSINR} with $E[P^{\frac{2}{\alpha}}]$
and $f_P(p)$ from Lemma \ref{Lemma:ErgodicSpatialAverage}.
Removing the conditioning with respect to $r_1$ and $\Pi_t$
yields the following property of the area-averaged spectral
efficiency $E[C]$ (where the averaging is over the fading and
wireless-node locations, base-station locations and
representative link length), as $n,N, R\to\infty$
\begin{align}
E[C] - \int\log_2(1+ P_1 r_1^{-\alpha}N^{\alpha/2}\beta)f_{r_1}(r1)\,dr_1 \to 0, \label{Eqn:UnCondMSEPoisson}
\end{align}
where we have used the fact that $\beta$ equals the same value
over almost all realizations of $\Pi_t$, and the monotone
convergence theorem (see e.g. \cite{Karr}) to exchange the
limit and expectation.

Although $\beta$ has to be found numerically in general, by
assuming that $c = n/N$ is large, as done in Section III, the
area-averaged spectral efficiency conditioned on $r_1$ is
approximated from \eqref{Eqn:CondMSEPoisson} as follows
\begin{align}
E[C|\Pi_t, r_1]
\approx \log_2\!\left(\!1+G_\alpha P_1 \!\left(\frac{N}{E\!\left[P^{\frac{2}{\alpha}} \right]\pi \rho_w r_1^2}\!\right)^{\frac{\alpha}{2}}\right)\, \label{Eqn:CondMeanSpecEff}
 \end{align}
which holds with probability 1. Furthermore, if
$\frac{p_t}{G_t}r_{1}^\alpha < P_M$, i.e., the transmit power
budget is sufficient for the representative transmitter to
achieve the target received power $p_t$ at the representative base station,
substituting \eqref{Eqn:PowerControlHex} and
\eqref{Eqn:TetheredNodePowerScale} into
\eqref{Eqn:CondMeanSpecEff} yields the following approximation which holds with probability 1.
\begin{align}
&E[C|\Pi_t, r_1] \approx\nonumber \\
& \log_2\!\left(\!1+G_\alpha  \!\left(\frac{\rho_t\,N}{\left(1-e^{-\pi\rho_t{\left(\frac{G_t}{p_t}P_M\right)}^{\frac{2}{\alpha}}}\right)\!\rho_w}\!\right)^{\frac{\alpha}{2}}\right) \label{Eqn:CondMeanSpecEff2}
 \end{align}
With probability 1, the above expression approximates the
area-averaged spectral efficiency of a wireless link that has a
sufficient power budget to meet its target received power where
the average is taken over the fading and wireless-node
distributions. If we further assume that the transmit power
budget $P_M$ is large, the exponential term in the expression
above is small, resulting in the following simple expression
for the area-averaged spectral efficiency which holds with
probability 1.
\begin{align}
E[C|\Pi_t] &\approx \log_2\left(1+ G_\alpha \left(\frac{N\rho_t}{\rho_w}\right)^{\frac{\alpha}{2}}\right) \approx E[C]. \label{Eqn:MeanSpecEffTetheredNodesNoBudget}
\end{align}
Note that the approximation above holds with probability 1 and
is essentially not dependent on the specific realization of
$\Pi_t$ as a consequence of the ergodicity of the PPP and the
large number of degrees of freedom at the MMSE receiver, which
makes the system less sensitive to variations in the
base-station positions. Additionally, note that the
area-averaged spectral efficiency from
\eqref{Eqn:MeanSpecEffTetheredNodesNoBudget} primarily depends
on $\rho_t/\rho_w$,implying approximate scale invariance in networks where
the power budget $P_M$ is not a significant limitation. The
scale invariance indicates that as with hexagonal cells (from
equation \eqref{Eqn:HexSufficientPowerByDensity}),
approximately constant area-averaged spectral efficiency can be
maintained by fixing the relative density of base stations to
interferers.
\begin{figure*}[!t]

\normalsize
\setcounter{MYtempeqncnt}{\value{equation}}
\setcounter{equation}{31}
\begin{align}
&E[C] \approx \int_0^\infty\log_2\left(1+\min\left(\frac{p_t}{Gt}r_1^\alpha, P_M\right)r_1^{-\alpha}G_\alpha \left(\frac{N}{E[P^{\frac{2}{\alpha}}]\pi\rho_w}\right)^{\frac{\alpha}{2}}\right)  2\pi\rho_t r_1 e^{-\pi\rho_tr_1^2} dr_1 =\left( 1-e^{-\pi\rho_t{\left(\frac{G_t}{p_t}P_M\right)}^{\frac{2}{\alpha}}} \right)\nonumber\\
&\times \log_2\left(1+\frac{p_t}{G_t} G_\alpha \left(\frac{N}{E[P^{\frac{2}{\alpha}}]\pi\rho_w}\right)^{\frac{\alpha}{2}}\right) +  \int_{r_t}^\infty\log_2\left(1+P_Mr_1^{-\alpha} G_\alpha \left(\frac{N}{E[P^{\frac{2}{\alpha}}]\pi\rho_w}\right)^{\frac{\alpha}{2}}\right)  2\pi\rho_t r_1 e^{-\pi\rho_tr_1^2} dr_1. \label{Eqn:MeanSpecEffTetheredNodes}
\end{align}
\vspace*{4pt}
\hrulefill
\setcounter{equation}{\value{MYtempeqncnt}}
\addtocounter{equation}{1}
\end{figure*}
If the transmit power budgets $P_M$ are not sufficiently large
to permit the approximations in
\eqref{Eqn:MeanSpecEffTetheredNodesNoBudget}, we can use
\eqref{Eqn:UnCondMSEPoisson} with the approximation for $\beta$
and by observing that $r_1$ follows the nearest neighbor
distribution for Poisson point processes \cite{Stoyan} as
follows
\begin{align}
f_{r_{1}}(r_1) = 2\pi\rho_t r_1 e^{-\pi\,\rho_t \, r_1^2}\,\mbox{ for } r_1 > 0\,. \label{Eqn:NearestNeighborPDF}
\end{align}
The area-averaged spectral efficiency of a random link averaged
over realizations of $\Pi_t$ can then be found by removing the
conditioning on $r_1$ and $\Pi_t$ by substituting
\eqref{Eqn:PowerControlHex} into \eqref{Eqn:CondMeanSpecEff}
and integrating with respect to the density in
\eqref{Eqn:NearestNeighborPDF} which yields \eqref{Eqn:MeanSpecEffTetheredNodes} at the top of this page.
We were unable to find a closed form expression for the second
term on the RHS of  \eqref{Eqn:MeanSpecEffTetheredNodes} and
thus use numerical integration to evaluate it.

\subsection{Comparison Between Random and Hexagonal Cells}
\begin{figure}
\begin{center}
\includegraphics[width=3.75in]{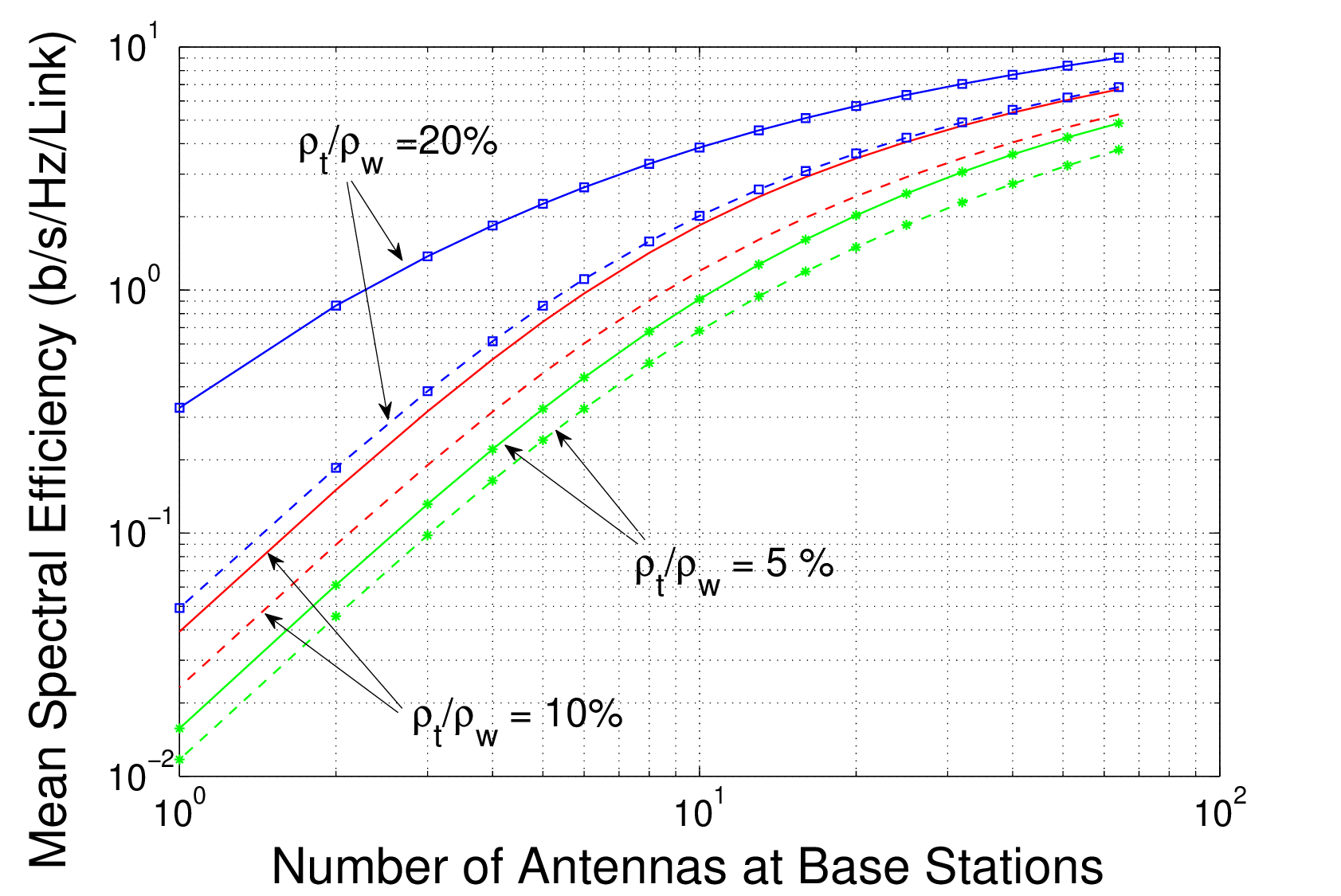}
\caption{ \label{Fig:HexVsRandom}
Area-averaged spectral efficiency of the uplink  with random cells and hexagonal cells and transmit power limited to 200 mW. Solid and dashed lines  represent hexagonal and random cell asymptotic spectral efficiencies respectively. $\rho_t$ and $\rho_w$ are the base station and wireless node density respectively.}
\end{center}
\end{figure}

For
systems with limited transmit powers, we numerically evaluated
and plotted equations for the spectral efficiency corresponding
to random and hexagonal cells in  Figure \ref{Fig:HexVsRandom},
where the solid and dashed lines represent hexagonal and random
cells respectively. The transmit power budget was 200 mW and
wireless-node density was $10^{-3}$ with different relative
density of base stations to active interferers as shown in the
plot. Note that the difference in area-averaged spectral
efficiencies diminishes with the number of antennas. However,
for high base-station densities the area-averaged spectral
efficiency for random cells is significantly lower. For
instance, with 10 antennas at the base stations and 20\%
relative density of base stations to wireless nodes, the
area-averaged spectral efficiency with hexagonal cells is twice
that of random cells.

When compared to the area-averaged spectral efficiency with
random cells given by
\eqref{Eqn:MeanSpecEffTetheredNodesNoBudget},
\eqref{Eqn:HexSufficientPowerByDensity} indicates that
several-fold (but not orders of magnitude) gains in
area-averaged spectral efficiency can be achieved by regularly
distributing base stations in planar networks compared to
randomly distributing them, and furthermore, the difference
diminishes with the number of base-station antennas. In
practical systems, designers will of course not have the
flexibility of placing base stations and mobile user
distributions are not uniformly random (i.e. without spatial
correlations). Nevertheless, this result sheds some light into
the performance differences between these two idealized models
which are commonly used in the research community.

\section{Monte Carlo Simulations}  \label{Sec:MonteCarlo}

\subsection{Hexagonal Cells} \label{Sec:MonteCarloHex}

To verify the asymptotic results of the previous section, we
simulated network topologies with base stations at hexagonal
lattice sites, and interferers distributed randomly on a large
circular network on the plane. We simulated each configuration
5000 times. The representative transmitter was placed with
uniform probability  in the center-most cell.

For each trial, we placed 4000  interferers randomly in
circular networks with radii  selected to meet target
wireless-node  densities of $10^{-2} , 10^{-3}$, and $ 10^{-4}$
nodes $m^{-2}$. The circular network was overlayed on a
hexagonal grid of base stations which extends beyond the edge
of the circular network of interferers. The base stations were
spaced such that their densities were $20\%$, $10\%$, $5\%$ and
$2.5\%$ of the wireless-node density. We simulated systems with
both unlimited transmit powers (to simulate the
sufficient-power case) and powers limited to $P_{M} = 200mW$.

The channel coefficient between the antenna of  wireless node
$i$ and antenna $j$ of the representative  base station was
modeled as $\sqrt{G_tr_{i}^{-\alpha}}g_{ij}$, where $\alpha =
4$, $G_t = 10^{-5}\,m^4$, and $g_{ij}$ are i.i.d.
$\mathcal{CN}(0,1)$ random variables which represents the
narrow-band Rayleigh fading channel.

\subsubsection{Sufficient Transmit Powers}

\begin{figure}
\begin{center}
\includegraphics[width=95mm]{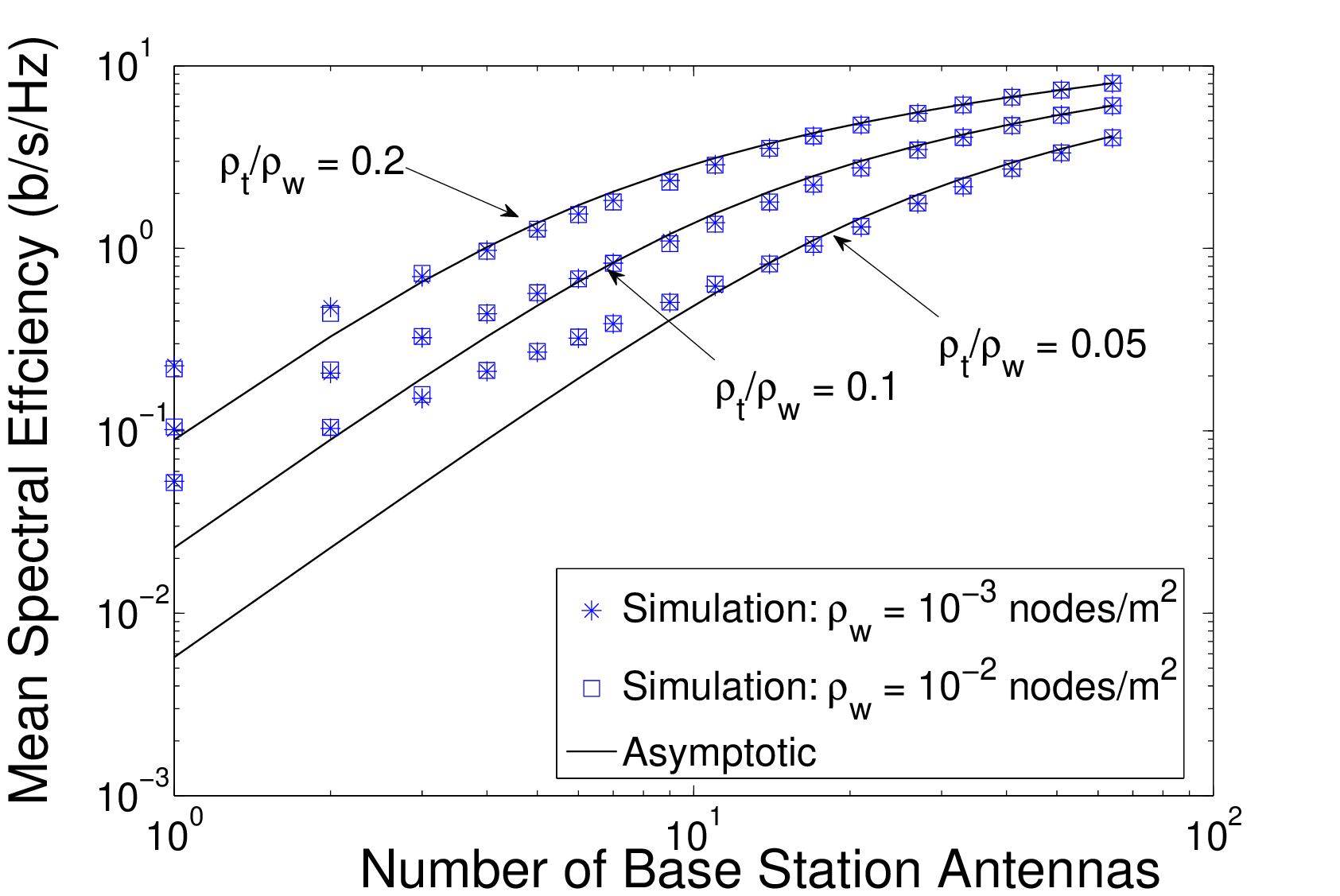}
\caption{Area-averaged spectral efficiency vs. number of receive antennas for  wireless-node densities of $\rho_w =  10^{-3}$ and $\rho_w = 10^{-2}$ nodes $m^{-2}$ with unlimited transmit powers and hexagonal cells with base-station density of $\rho_t$. \label{Fig:HexTetheredMeanSpecEff0_0001Target30dBUnlPower}}
\end{center}
\end{figure}

Figure \ref{Fig:HexTetheredMeanSpecEff0_0001Target30dBUnlPower}
illustrates the area-averaged uplink spectral  efficiency for
wireless-node densities of $\rho_w = 10^{-3}$ and $\rho_w =
10^{-2}$ nodes $m^{-2}$, and unlimited transmit powers per node
versus the number of antennas at the representative base
station.  The square and asterisk markers represent
wireless-node densities of $ 10^{-2}$, and   $ 10^{-3}$ nodes
$m^{-2}$, respectively and the solid lines represent the
asymptotic area-averaged spectral efficiency from
\eqref{Eqn:HexSufficientPower}.

Note that the asterisk and square markers coincide indicating
that the absolute  density of interferers does not effect the
area-averaged spectral efficiency appreciably, and it is the
relative density of interferers to base stations that matters.
Furthermore, note  that the asymptotic approximation
\eqref{Eqn:HexNodesInsuffPowerMean} holds when $N$ is
sufficiently large. When the base-station density is 20\% of
the \emph{active} wireless-node density, the asymptotic and
simulated area-averaged spectral efficiency differ by less than
10\% when $N\geq 10$. For lower densities of base stations, the
convergence is slower, e.g. when the base-station density is
5\% of the active wireless-node density, the difference between
the simulated and asymptotic area-averaged spectral efficiency
drops below 10\% only when $N > 37$.

\begin{figure}
\begin{center}
\includegraphics[width=95mm]{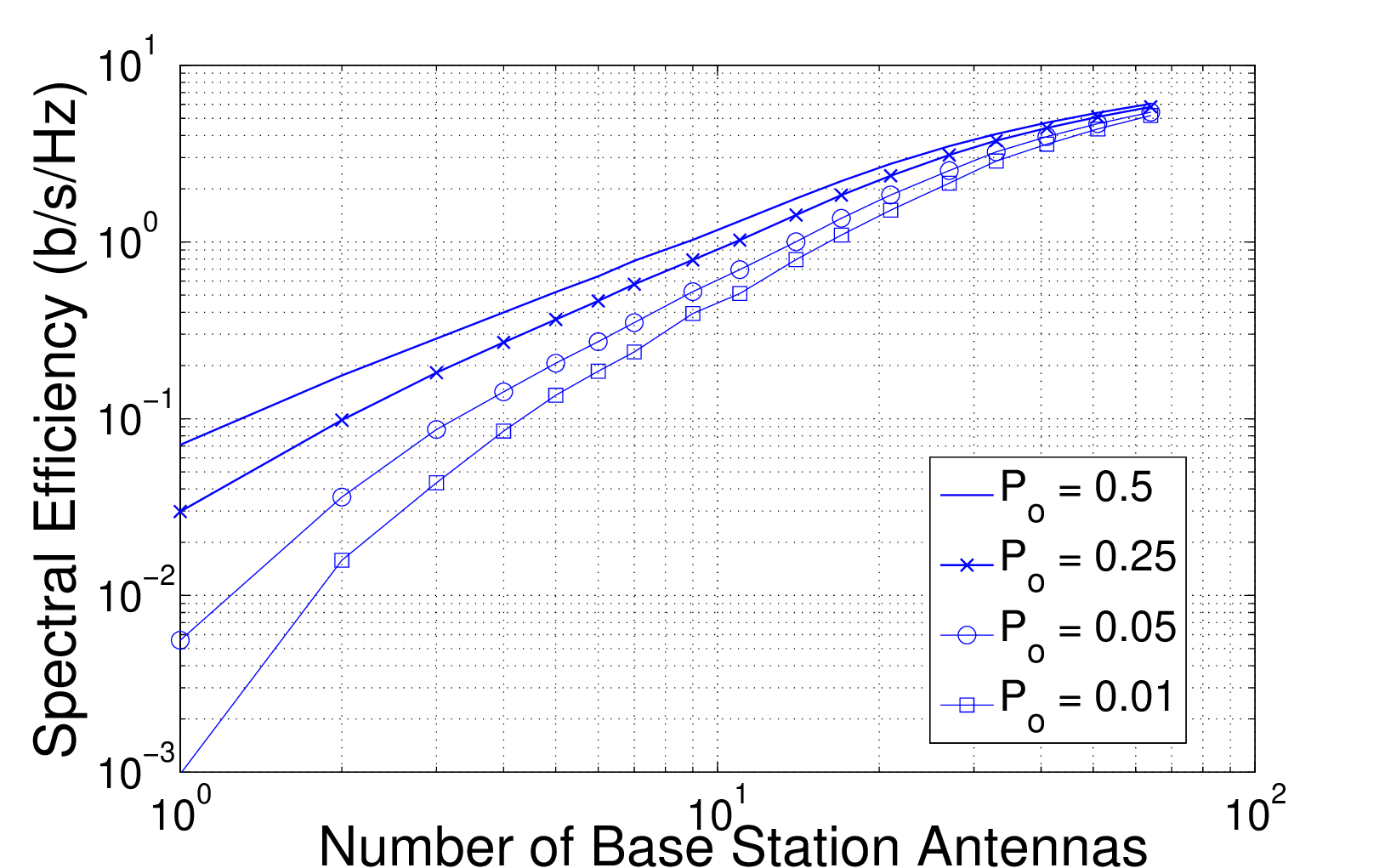}
\caption{Outage spectral efficiency vs. number of receive antennas for wireless-node densities of  $\rho_w = 10^{-2}$ nodes $m^{-2}$ with unlimited transmit powers and base-station density equaling $10\%$ of wireless-node density, with hexagonal cells. \label{Fig:HexTetheredOutSpecEffTarget30dBUnlPower01Density}}
\end{center}
\end{figure}

We analyzed the outage spectral efficiencies from the simulated
data, where spectral efficiency with outage probability  $P_o$
means that a fraction $1-P_o$ of the links in our simulations
achieved that spectral efficiency or greater. Figure
\ref{Fig:HexTetheredOutSpecEffTarget30dBUnlPower01Density}
illustrates the outage spectral efficiencies vs. number of
receive antennas on the representative link  for $\rho_w =
10^{-2}$ nodes $m^{-2}$ with $5\%$, $25\%$ and $50\%$ outage
probabilities. Note that the intersection of the line with the
circular markers and the 1 $b s^{-1} Hz^{-1}$ mark in Figure
\ref{Fig:HexTetheredOutSpecEffTarget30dBUnlPower01Density}
occurs  approximately at $N = 14$ indicating that it is
possible for $95\%$ of links to achieve  1 $b s^{-1} Hz^{-1}$
with $N \geq 14$ when the base-station density is 10\% of the
density of  \emph{transmitting} interferers. In real systems,
the number of nodes transmitting at any time is far smaller
than the total number of nodes in the network. Suppose that at
any one time, $10\%$ of nodes are actively transmitting in the
network. Figure
\ref{Fig:HexTetheredOutSpecEffTarget30dBUnlPower01Density}
indicates that with a base-station density equaling $1\%$ of
total wireless-node density (including inactive ones), it is
possible for $95\%$ of links to achieve  1 $b s^{-1} Hz^{-1}$
with 14 antennas at each base station.

\subsubsection{Insufficient Transmit Power}

\begin{figure}
\begin{center}
\includegraphics[width=95mm]{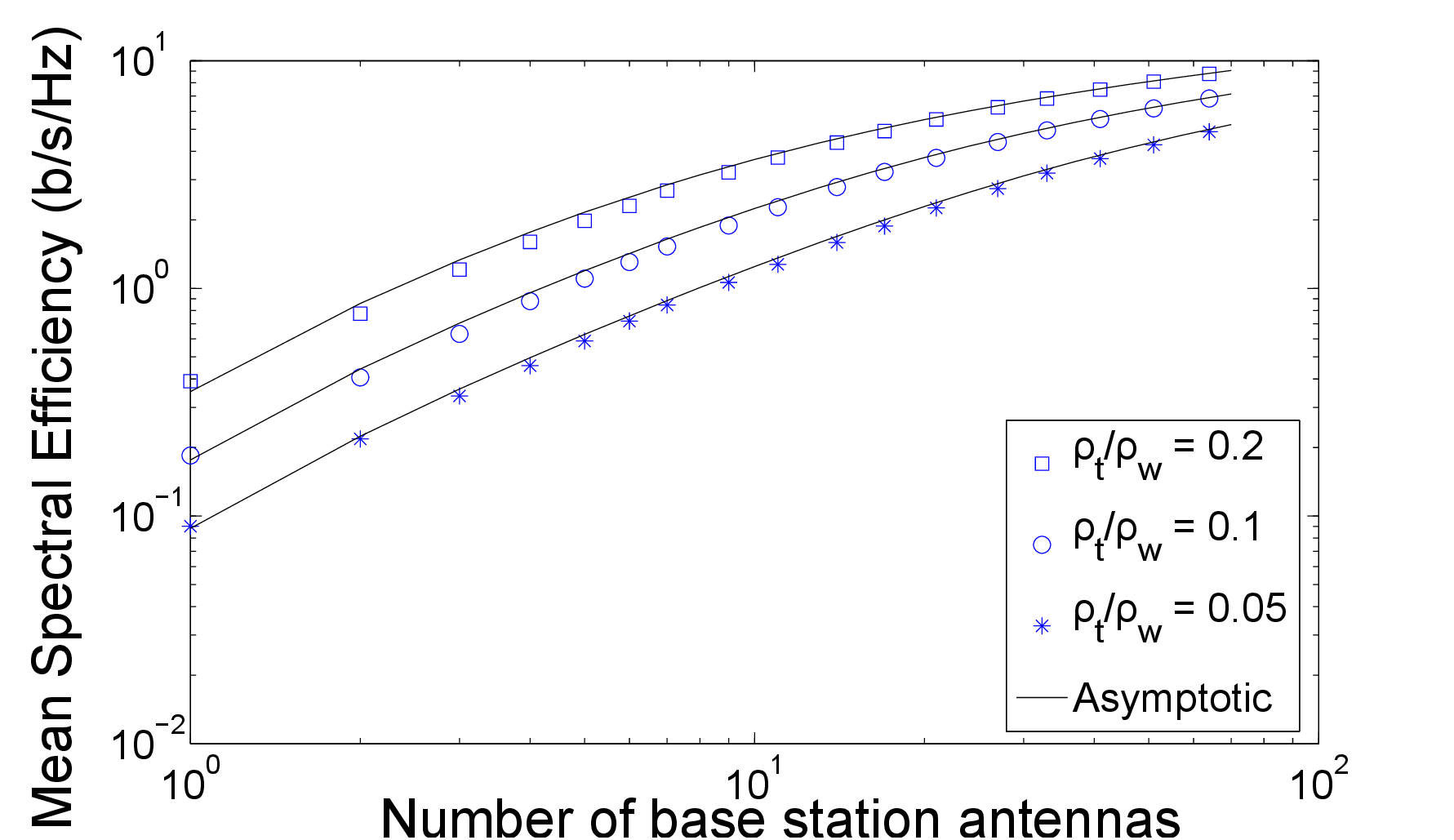}
\caption{Area-averaged spectral efficiency for $\rho_w =  10^{-4}$ nodes $m^{-2}$ with different relative density of base station to interferers and hexagonal cells, and 200mW transmit power limits per wireless node. \label{Fig:HexTetheredMeanSpecEff0_0001Target30dB}}
\end{center}
\end{figure}

\begin{figure}
\begin{center}
\includegraphics[width=95mm]{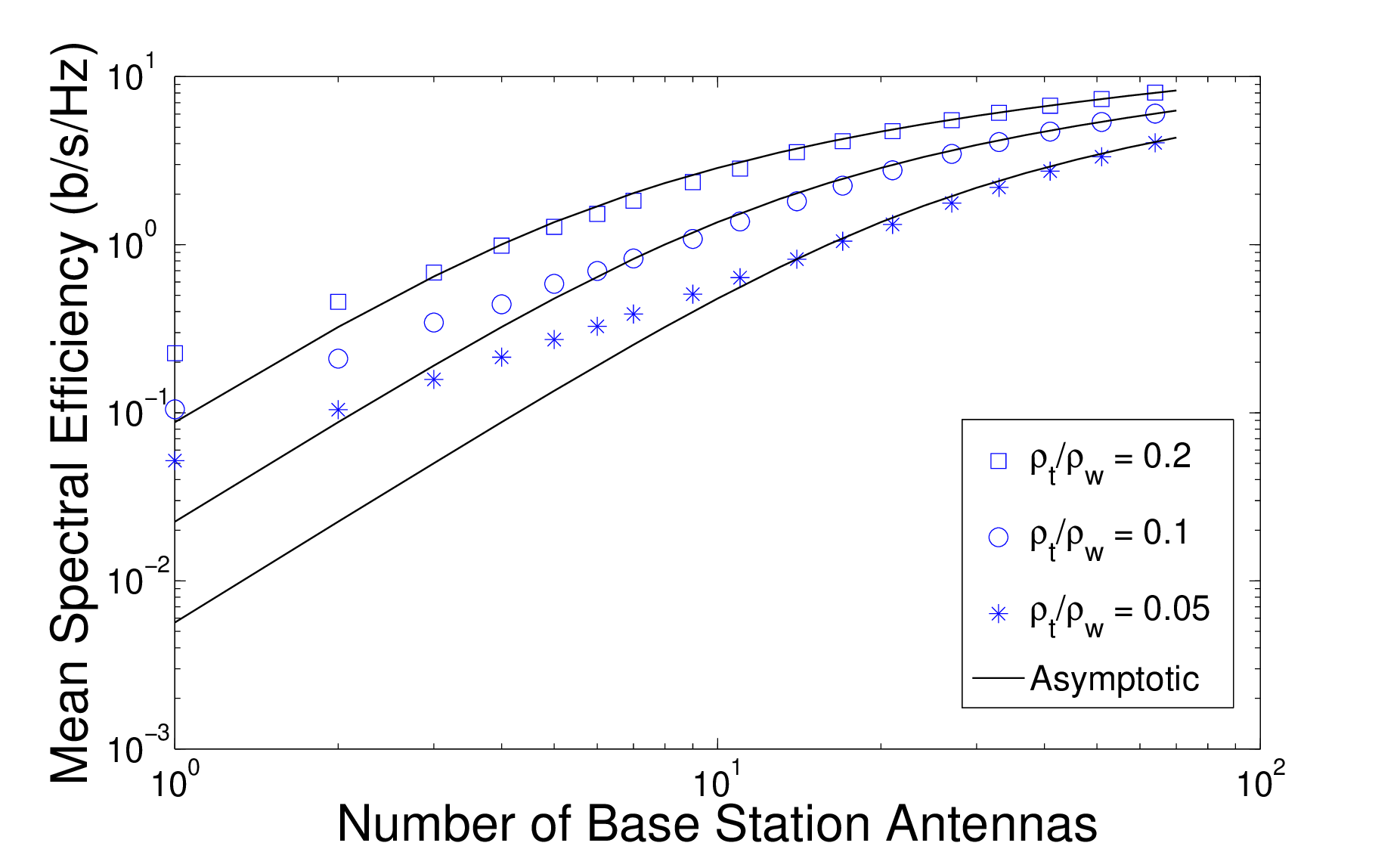}
\caption{Area-averaged spectral efficiency vs. number of receive antennas for $\rho_w = 10^{-2}$ nodes $m^{-2}$ with different relative density of base stations to interferers and hexagonal cells, and 200mW transmit power limits per wireless node.  \label{Fig:HexTetheredMeanSpecEff0_01Target30dB}}
\end{center}
\end{figure}

Figures \ref{Fig:HexTetheredMeanSpecEff0_0001Target30dB} and
\ref{Fig:HexTetheredMeanSpecEff0_01Target30dB} illustrate the
area-averaged spectral efficiency vs. number of receive
antennas for $\rho_w = 10^{-4}$ and $\rho_w = 10^{-2}$
respectively, with $P_M = 200 mW$. The  markers represent the
simulated area-averaged spectral efficiencies for different
relative densities of base stations to interferers.  The solid
lines are the predicted asymptotic area-averaged spectral
efficiencies obtained by numerically evaluating equation
\eqref{Eqn:HexNodesInsuffPowerMean}.

It is clear from Figures
\ref{Fig:HexTetheredMeanSpecEff0_0001Target30dB} and
\ref{Fig:HexTetheredMeanSpecEff0_01Target30dB} that the
asymptotic approximation  \eqref{Eqn:HexNodesInsuffPowerMean}
holds when $N$ is sufficiently large. In Figure
\ref{Fig:HexTetheredMeanSpecEff0_0001Target30dB}, the simulated
and asymptotic area-averaged spectral efficiencies agree to
within 5\% for $N \geq 2$ for all the base-station densities
considered. In Figure
\ref{Fig:HexTetheredMeanSpecEff0_01Target30dB} however, for
base-station densities that are 5\% of the wireless-node
density of $10^{-2}$ nodes $m^{-2}$, the simulated and
asymptotic spectral efficiencies differ by less than 13\% only
when there are 13 or more antenna elements at the receiver. For
base-station densities that are 20\% of the wireless-node
density, the simulated and asymptotic spectral efficiencies
agree to within 13\% when $N \geq 3$.

At low wireless-node densities, the simulated spectral
efficiencies  converge more  rapidly (compared to high
densities) to the asymptote because  a larger fraction of nodes
transmit at the power limit. The e.d.f. of interference powers
at the representative receiver thus converges more rapidly to
its asymptotic value. The rate of convergence of the e.d.f.  of
interference powers controls the rate of convergence of the
eigenvalues of the spatial interference covariance matrix
$\sum_{i = 2}^{n+1} p_i\,\mathbf{g}_i\,\mathbf{g}_i^\dagger$
(see Appendix \ref{Sec:MainTheoremProof} and Section 3 of
\cite{BaiSilversteinNoMinimum}) which affects the convergence
rates of the normalized SIR and spectral efficiency.

\begin{figure}
\begin{center}
\includegraphics[width=95mm]{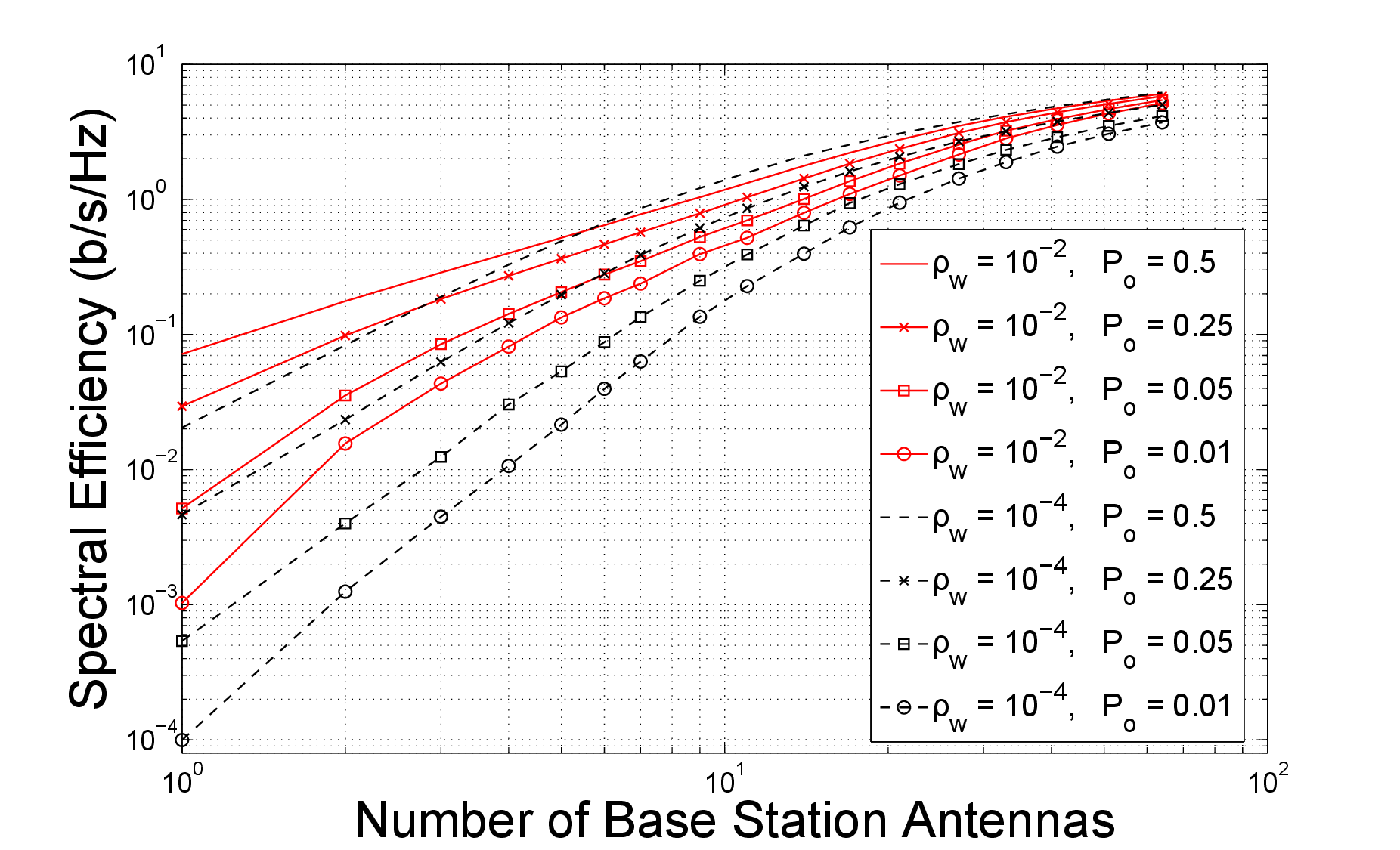}
\caption{Outage spectral efficiency vs. number of receive antennas for wireless-node density $\rho_w =  10^{-4}$ and $10^{-2}$ nodes $m^{-2}$ and base-station density equal to 10\% of wireless-node density, hexagonal cells and 200mW transmit power budget. The solid lines represent $\rho_w = 10^{-2}$ and dashed lines represent $\rho_w = 10^{-4}$. The markers represent the different outage probabilities, $P_o$ shown in the legend. \label{Fig:HexTetheredOutageSpecEff0_0001_0_001Target30dB}}
\end{center}
\end{figure}

Figure \ref{Fig:HexTetheredOutageSpecEff0_0001_0_001Target30dB}
shows the outage and area-averaged spectral efficiencies for
$\rho_w = 10^{-4}$ (solid lines) and  $\rho_w =  10^{-3}$
(dashed lines) nodes $m^{-2}$, $10 \%$ relative density of
base-stations to interferers and $P_M =  200 mW$. Note that
with 10 antennas at the receiver, area-averaged spectral
efficiencies of approximately 0.2 and 0.3 b/s/Hz are possible
for $\rho_w = 10^{-4}$ and $\rho_w = 10^{-3}$ respectively. The
discrepancy in the spectral efficiency is a result of the
maximum transmit power. For $\rho_w = 10^{-4}$, a larger
fraction of nodes transmit at $P_M$ compared to $\rho_w =
10^{-3}$, resulting in higher SIRs for $\rho_w = 10^{-4}$. The
higher total interference power for $\rho_w = 10^{-3}$ is
offset by increased signal powers due to shorter links since
the relative base-station to wireless-node density is fixed.

\subsection{Poisson Cell Model}
We verified  \eqref{Eqn:MeanSpecEffTetheredNodes} and
\eqref{Eqn:MeanSpecEffTetheredNodesNoBudget} by Monte Carlo
simulations of the network topology. We placed  base stations
in a circular network of radius 4$R$. The numbers of base
stations were selected to achieve relative densities of base
stations to interferers of 20\%, 10\% and 5\%. The network of
base stations was then re-centered such that a base-station is
at the origin.   4000 interferers were then placed in a
circular network of radius $R$, centered on the base station at
the origin with $R$ selected to achieve a wireless-node density
of $10^{-3}$ nodes m$^{-2}$.  This experiment was repeated 1000
times. For each trial, the spectral efficiency of a link placed
in the center-most cell  was collected and averaged. The
transmit power of each wireless node was set according to
\eqref{Eqn:PowerControlHex} with $P_M = \infty$ (to simulate
the sufficient power case) or $P_M = 200\, mW$. $G_t =
10^{-5}m^\alpha$,  and $\alpha = 4$, were assumed.

\begin{figure}
\begin{center}
  \includegraphics[width=95mm]{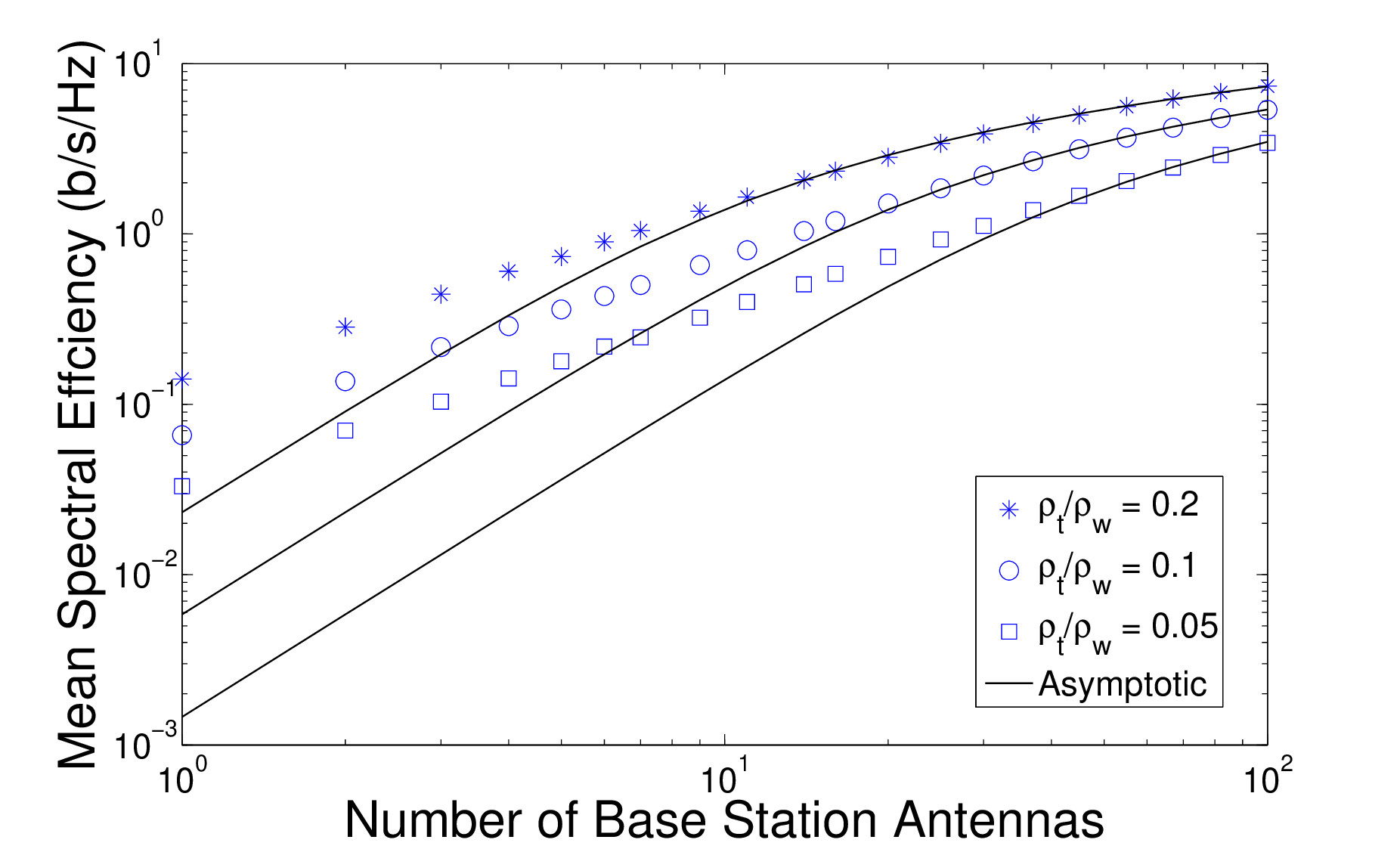}
\caption{ \label{Fig:RandomTetheredNodesNoPowerLimit}
Area-averaged spectral efficiency of uplink communications with random cells and unlimited transmit powers. Wireless-node density $\rho_w = 10^{-3}$ nodes / $m^2$, $p_t = 10^{-14}$ and $G_t = 10^{-5}$.
}
\end{center}
\end{figure}

\begin{figure}
\begin{center}
\includegraphics[width=95mm]{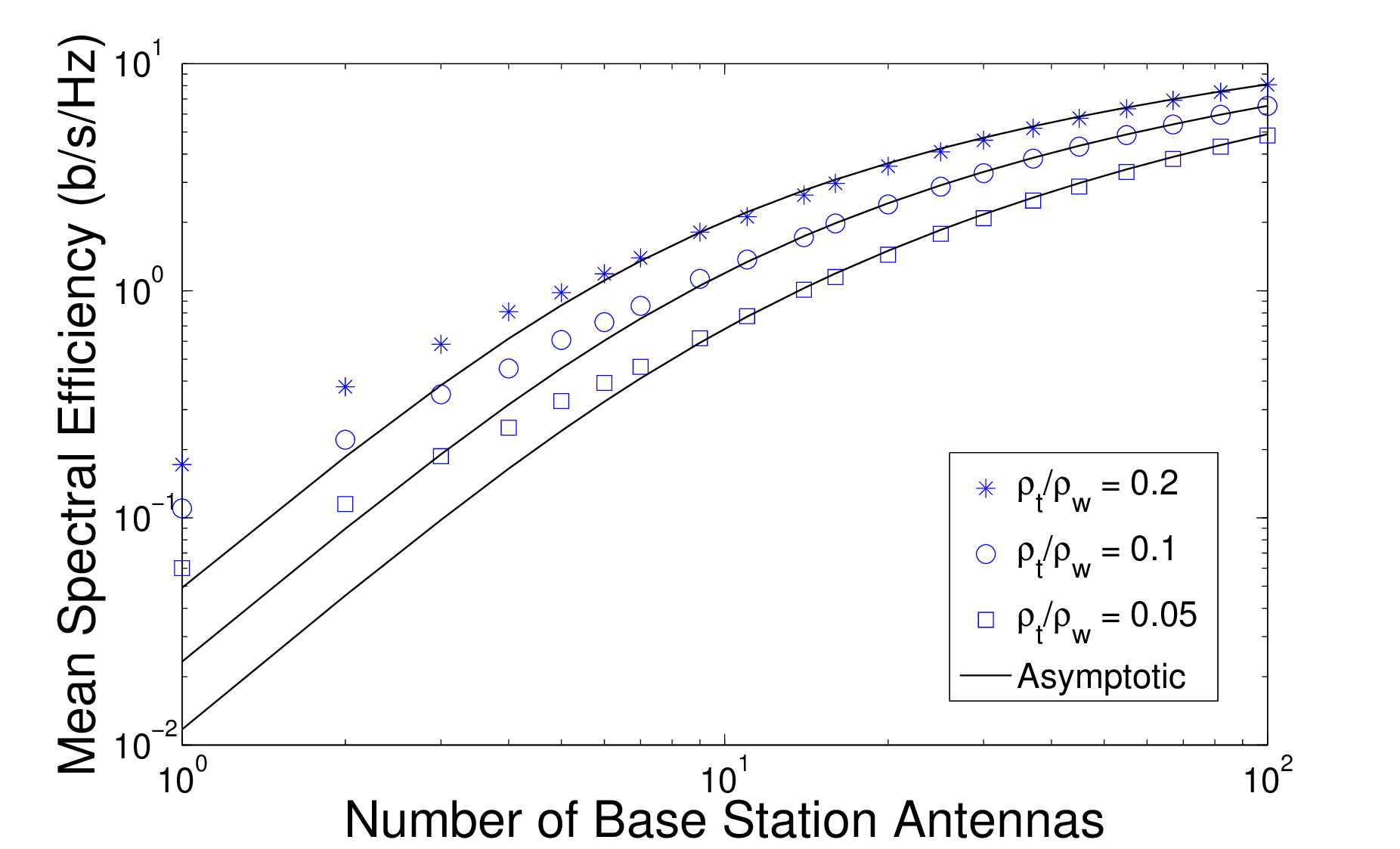}
\caption{ \label{Fig:RandomTetheredNodesPowerLimit}
Area-averaged spectral efficiency of uplink communications with random cells and 200mW transmit power limit per node. Wireless-node density $\rho_w = 10^{-3}$ nodes / $m^2$, $p_t = 10^{-14}$ and $G_t = 10^{-5}$.
}
\end{center}
\end{figure}

Figure \ref{Fig:RandomTetheredNodesNoPowerLimit} shows  results
of Monte Carlo simulations  and the asymptotic expression given
by \eqref{Eqn:MeanSpecEffTetheredNodesNoBudget} for systems
with unlimited transmit powers per node. Note that the
simulations match the asymptotic results to within 10\% when
$N\geq 9$ for a relative base-station to wireless-node density
of 20\%. For lower relative densities, the convergence is
slower. For 10\% relative density, the simulations match the
asymptotic expression to within 10\% only when $N \geq 20$ and
only when $N \geq 37$ for 5\% relative density. The rate of
convergence for random cells is slower than that for hexagonal
cells because the range of transmit powers is much larger for
random cells compared to hexagonal cells which results in
slower convergence, as explained in Section
\ref{Sec:MonteCarloHex}. Figure
\ref{Fig:RandomTetheredNodesPowerLimit} shows simulations of
systems with a 200 mW transmit power limit. The target received
power $p_t = 10^{-12}$.  For  relative base-station to
wireless-node densities of  20\%, 10\%, and 5\%, the simulated
area-averaged spectral efficiencies are within 10\% of the
asymptotic prediction when $N \geq 6$,  $N\geq 9$ and $N \geq
9$ respectively. The convergence of the simulated area-averaged
spectral efficiencies to the asymptotic values is faster for
systems with limited transmit power as the range of transmit
powers in the network is smaller when there is a bound on the
transmit power.

\section{Summary and Conclusions}
We have derived an asymptotic expression for the area-averaged
spectral efficiency of the uplink in wireless networks with
multi-antenna base-stations in networks with hexagonal cells, a
model which is known to be difficult to analyze and is
typically handled in simulation. We extended the results to
networks where base stations are distributed according to a
Poisson point process on the plane and derive an expression for
the area-averaged spectral efficiency of a random link where
the averaging is  over the fading, and spatial distributions of
the wireless nodes and base stations. We assumed a power
control algorithm for which interferers try to achieve a target
received power at the base stations to which they are
connected. This  power control algorithm which has also been
used in \cite{AndrewsOutage} and related works ensures that
uplink spectral efficiencies are close to the average value
with high probability when the number of antennas per link is
large and the interferers have high power budgets.

It is found that if the cell sizes are small enough that all
interferers are able to achieve the target received signal
power at their base stations (which we call the
sufficient-power case), the area-averaged spectral efficiency
takes a simple form given by \eqref{Eqn:HexSufficientPower}.
Note that for a fixed ratio of base-station to wireless-node
densities $\rho_t/\rho_w$, as $\rho_w$ increases, the system
eventually moves to the sufficient-power case so this is an
effective way of scaling the density of such networks.

From \eqref{Eqn:HexSufficientPower}, note that with 7 antenna
elements per base station and $\rho_t/\rho_w \approx 0.1$, the
area-averaged spectral efficiency is approximately 1 $b
s^{-1}Hz^{-1}$. If we assume that 10\% of all interferers are
actively transmitting at any one time,  the ratio of  base
station to total wireless-node density has to be just $1\%$ to
achieve a area-averaged spectral efficiency of 1 $b
s^{-1}Hz^{-1}$, as given by \eqref{Eqn:HexSufficientPower}. For
systems with insufficient power, i.e., the cells are so large
that not all of the interferers will achieve the target
received power at their base stations, the expression for the
area-averaged spectral efficiency is more complicated and has
to be evaluated by numerically. We verified the accuracy of the
derived expressions by Monte Carlo simulations which were also
used to study the outage spectral efficiency, i.e., the
spectral efficiency that is achievable with a given
probability. We found that in the sufficient power case, with
14 antennas per base station and single antennas at each
wireless-node, and with 10\% of interferers \emph{transmitting
simultaneously} at any one time, over 1 $b s^{-1} Hz^{-1}$ is
achievable by 95\% of interferers when the ratio of
base-station to active wireless-node densities is $1\%$.

Comparing the area-averaged spectral efficiencies for hexagonal
and Poisson cells, we find that the difference in area-averaged
spectral efficiency between the two models diminishes with
increasing $N$. At modest $N$ we found that hexagonal cells can
increase the area-averaged spectral efficiency over random
cells several-fold as illustrated in Figure
\ref{Fig:HexVsRandom}.

The findings of this work are useful for designers of cellular
wireless systems such as pico-cells and city-wide wi-fi access
as they provide simple expressions for the spectral efficiency
and hence data rates as a function of tangible system
parameters such as user and base-station densities,  number of
base-station antennas and random versus regular distribution of base stations.

\section{Acknowledgement}
We would like to thank the anonymous reviewers for their
comments which have greatly improved our exposition and the
positioning of our results with respect to existing literature.


\appendix

\subsection{Proof of Theorem 1} \label{Sec:MainTheoremProof}
To derive the normalized SIR $\beta_N$, we first modify the system model from Section \ref{Sec:NetworkModel}.
We assume that the
$i$-th wireless-node transmits with power $\tilde{P}_i =
N^{{\frac{\alpha}{2}}} P_i$ for $i = 2, 3, \cdots, n+1$, where
$P_i$ is as defined in Section \ref{Sec:NetworkModel}, whereas
the representative transmitter transmits with power $P_1$.
Thus, the SIR of this system , is equivalent to
$N^{-{\frac{\alpha}{2}}}$ times the SIR of the original system in
Section \ref{Sec:NetworkModel} where the interferers transmit
with power $P_i$. Let the matrix $\mathbf{P} =
diag(\tilde{P}_2r_2^{-\alpha},\tilde{P}_3r_3^{-\alpha}, \cdots,
\tilde{P}_{n+1}r_{n+1}^{-\alpha})$. The SIR of this system
normalized by $p_1$ is
\begin{align}
\beta_N &= \frac{1}{N}\mathbf{g}_1^\dagger
\left(\frac1N\mathbf{G}\mathbf{P}\mathbf{G}^\dagger\right)^{-1}\mathbf{g}_1 \label{Eqn:NormSIRDef}
\end{align}
where $\mathbf{G}$ is a matrix whose $i$-th column is
$\mathbf{g}_{i+1}$. Note that \eqref{Eqn:SIROrigDef} and
\eqref{Eqn:NormSIRDef} are equal.

Observe that \eqref{Eqn:NormSIRDef} and
\eqref{Eqn:NormSIRDefGenLemma} in  Lemma
\ref{Lemma:SIRConvergenceLemma} take the same form if the
e.d.f. of the diagonal entries of $\mathbf{P}$ converges $w. p.
1$ to a limiting function $H(x)$ as $N$ and $n \to \infty$, and
there exists an $N_0$ such that for all $N > N_0$, the
eigenvalues of $\frac1N\mathbf{GPG}^\dagger$ are bounded from
below. The latter requirement is satisfied as shown in the
following lemma.

\begin{lemma} \label{Lemma:MinEval}
Let $\lambda_{min}(\mathbf{A})$ denote the minimum eigenvalue
of the matrix $\mathbf{A}$. Consider the matrix
\begin{align}
\mathbf{K} = \frac1N\mathbf{GPG}^\dagger = N^{{\frac{\alpha}{2}}-1} \sum_{i= 2}^{n+1}
p_i\mathbf{g}_i\mathbf{g}_i^\dagger\,.
\end{align}
Then $ 0 < \lambda_{\ell b} < \lambda_{min}(\mathbf{K}), \; w.
p. 1$ for some $\lambda_{\ell b}$ and $\forall n
> N_0$ where $N_0$ is a positive integer.

\end{lemma}
\noindent{\it Proof:} Please see Appendix
\ref{Lemma:MinEvalProof}

Hence, what remains is to show the convergence of the e.d.f. of
the received interference powers. Recall that $n$ interferers
are distributed in a disk of radius $R$ centered at the origin.
Setting $G_t = 1$ for notational convenience (it will be
reintroduced in the final expressions) and $\tilde{p}_i =
\tilde{P}_ir_i^{-\alpha}$, the CDF of the received power from
wireless-node $i$ is
\begin{align}
&\Pr\{\tilde{p}_i \leq x\} =  \Pr\{P_iN^{\frac{\alpha}{2}}r_i^{-\alpha} \leq x\} \nonumber \\
&\;\;\;\;\;\;\;= \int \Pr\left\{\frac{r_i}{\sqrt{N}} \geq \left. \left(\frac{P_i}{x}\right)^{\frac{1}{\alpha}} \right| P_i\right\} f^{N}_P(P_i)dP_i  \,, \label{Eqn:PowerCondCDF}
\end{align}
where $f^{N}_P(P_i)$ is the PDF of $P_i$ for a finite $N$.

Next, we use the following lemma which is proved in Appendix
\ref{Sec:CellMildProof}
\begin{lemma}\label{Lemma:CellMild}
For both the hexagonal-cell and Poisson-cell models, as $n, N,
R \to\infty$ ,
\begin{align}
&\Pr\left\{\frac{r_{i}}{\sqrt{N}} \geq  \left(\frac{P_i}{x}\right)^{\frac{1}{\alpha}},P_i \right\} \to  \nonumber \\ &\;\;\;\;\;\;\left[\left(1-\frac{\pi\rho_w}{c}\left(\frac{P_i}{x}\right)^{\frac{2}{\alpha}} \right)I_{ \left\{ P_i\left(\frac{\pi\rho_w}{c}\right)^{{\frac{\alpha}{2}}} < x \right\}} \right] f_P(P_i)\,. \label{Eqn:PowerDistRest}
\end{align}
\end{lemma}
\eqref{Eqn:PowerDistRest} indicates that the transmit power
of a node randomly distributed with uniform probability in the
circular network to be  asymptotically independent of its
normalized distance from the origin as the quantity in the
brackets in \eqref{Eqn:PowerDistRest} equals the probability
that $r_i/\sqrt{N}$ exceeds
$\left(P_i/x\right)^{\frac{1}{\alpha}}$. From
\eqref{Eqn:PowerCondCDF}, Lemma \ref{Lemma:CellMild}, and the bounded convergence theorem,  as $n,
N, R\to \infty$ in the manner of Lemma \ref{Lemma:CellMild},
\begin{align}
&\Pr\{\tilde{p}_i \leq x\}\to \int f_P(P)\left(1-\frac{\pi\rho_w}{c}\left(\frac{P}{x}\right)^{\frac{2}{\alpha}}\right)I_{  \left\{P < \frac{x}{b} \right\}}\,dP  \label{Eqn:RxPowNodeDist}\\
&= F_P\left(\frac{x}{b}\right)-\frac{\pi\rho_w}{c}x^{-\frac{2}{\alpha}}E\left[P^{{\frac{2}{\alpha}}}\right]\nonumber \\ &\;\;\;\;\;\;\;\;\;\;\;\;\;\;\;\;\;\;\;\;+\frac{\pi\rho_w}{c}x^{-\frac{2}{\alpha}}\int_{\frac{x}{b}}^\infty f_P(P)P^{\frac{2}{\alpha}}\,dP\,. \label{Eqn:LimitingEDFBasic}
\end{align}
\eqref{Eqn:PowerCondCDF} to \eqref{Eqn:RxPowNodeDist} follows
from \eqref{Eqn:PowerDistRest}, and from substituting
$b=\left(\frac{\pi\rho_w}{c}\right)^{{\frac{\alpha}{2}}}$.

By the Glivenko-Cantelli theorem, the e.d.f. of a set of i.i.d.
random variables converges uniformly, $w. p. 1$, to its CDF.
The deviation of this e.d.f. from the CDF can be bounded by an
exponentially decreasing function of $n$, \emph{independent of the
CDF} \cite{dvoretzky}. Hence, by the Borel Cantelli Lemma, the e.d.f. of the $\tilde{p_i}$s converges $w. p. 1$ to
the RHS of \eqref{Eqn:LimitingEDFBasic}, i.e. $H(x) =
\Pr\{\tilde{p}_i < x \}$, even though the CDF is dependent on
$n$. Taking the derivative of the RHS of
\eqref{Eqn:LimitingEDFBasic} and simplifying yields:
\begin{align}
\frac{dH(x)}{dx} = \frac{2\pi\rho_w}{c\alpha}E&\left[P^{\frac{2}{\alpha}}\right]x^{-\frac{2}{\alpha}-1}  \nonumber \\
&-  \frac{2\pi\rho_w}{c\alpha}x^{-\frac{2}{\alpha}-1}\int_{x/b}^\infty f_P(\tau)\tau^{\frac{2}{\alpha}}d\tau.  \label{Lemma:DiffCorrelatedPowers}
\end{align}


Substituting \eqref{Lemma:DiffCorrelatedPowers} and integrating, the RHS of  \eqref{Eqn:FixedPoint}  becomes
\begin{align}
&m c \int_{b}^{\infty} \frac{\tau dH(\tau)}{1+\tau m} =   m c \int_0^\infty  \frac{2\pi\rho_w}{c\alpha}E\left[P^{\frac{2}{\alpha}}\right]\frac{\tau^{-\frac{2}{\alpha}}}{1+m\tau} d\tau \nonumber \\
&\;\;\;\;\;\;\;\;\;\;\;\;\;\;\;-  \frac{2\pi\rho_w m}{\alpha}  \int_0^\infty \frac{\tau^{-\frac{2}{\alpha}}}{1+m\tau}\, d\tau \, \int_{\tau/b}^\infty f_P(x)x^{\frac{2}{\alpha}}dx\nonumber \\
&=    \frac{2\pi\rho_w}{\alpha} E\left[P^{\frac{2}{\alpha}}\right] m^{\frac{2}{\alpha}} \pi\csc\left(\frac{2\pi}{\alpha}\right) \nonumber \\
&\;\;\;\;\;\;\;\;\;\;\;\;-  \frac{2\pi\rho_w m}{\alpha}  \int_0^\infty \frac{\tau^{-\frac{2}{\alpha}}}{1+m\tau}\, d\tau \, \int_{\tau/b}^\infty f_P(x)x^{\frac{2}{\alpha}}dx\,. \label{Eqn:TetheredNodesFixedPtIntegral1}
\end{align}
Substituting \eqref{Eqn:TetheredNodesFixedPtIntegral1} into \eqref{Eqn:FixedPoint} 
completes the proof.
\subsection{Proof of Lemma \ref{Lemma:SIRConvergenceLemma}}
\label{Lemma:SIRConvergenceLemmaProof} The proof mirrors that
of the proof of Lemma 4.3 in \cite{TseHanly} but since the
noise power is neglected in our model, a modified version of
the proof is required. Let $\lambda_1, \lambda_2, \cdots,
\lambda_N$ denote the eigenvalues of $\mathbf{R}$, and perform
an eigen decomposition of $\mathbf{R}$ such that $\mathbf{R} =
\mathbf{U}^\dagger\mathbf{\Lambda U}.$ Denoting the $i$-th
entry of the vector $\mathbf{Us}$ by $u_i$, we have
\begin{align}
 \gamma_N = \frac1N\sum_{i = 1}^{N}\frac{|u_i|^2}{\lambda_i}\,,
\end{align}
which is finite $w.p.1$ for $N > N_0$ since
$\lambda_{\ell\,b}<{\lambda_i}$. Note that as $n,N\to\infty$, the e.d.f. of the eigenvalues of $\mathbf{R}$, $\Psi_N(\lambda)$, converges with probability 1 to a limiting probability distribution $\Psi(\lambda)$ \cite{BaiSilverstein}.

For any $N > N_0$, set a
$\delta_1 > 0$ and pick a finite partition of the range
$(\lambda_{\ell \,b}, \infty)$ into $M$ intervals $(I_1, I_2,
\cdots I_M)$  such that
\begin{align}
&\sum_{k =1}^M  \frac{\Psi(I_k)}{I_k^l} - \int_0^\infty \frac{1}{\lambda}d\Psi(\lambda) < \delta_1\,, \;\;\;\;\;\mbox{and}
\\
&\int_0^\infty \frac{1}{\lambda}d\Psi(\lambda)  - \sum_{k =1}^M  \frac{\Psi(I_k)}{I_k^r} < \delta_1\,,
\end{align}
 where $\Psi(I_k)$ is the probability that a random variable with CDF $\Psi(\cdot)$ is in the interval $I_k$, and $I_k^l$ and $I_k^r$ are the left and right
edge of the interval $I_k$. Consider the events
\begin{align}
E_1 &= \left\{\left|\sum_{i:\lambda_i \in I_k} u_i^2 - \Psi_N(I_k) \right| < \frac{\delta_2}{M}, \forall \, k = 1, \cdots, M\right\} \;\; \mbox{and}\\
E_2 &= \left\{\left|\Psi_N(I_k) -\Psi(I_k) \right| < \frac{\delta_2}{M}, \forall \, k = 1, \cdots, M\right\}\, \label{Eqn:Event2Def}
\end{align}
where $\Psi_N(I_k)$ denotes the probability that a random variable with CDF $\Psi_N(\cdot)$  is in the interval $I_k$.
If both $E_1$ and $E_2$ hold, and $\sigma^2 = 0$,  following
\cite{TseHanly}, we have the following $w.p.1$.
\begin{align}
\gamma_N &\leq \sum_{k =1}^M \frac{\Psi(I_k) + 2\frac{\delta_2}{M}}{I_k^l} \leq \int_0^\infty \frac{1}{\lambda}d\Psi(\lambda) + \delta_1 + 2\frac{\delta_2}{\lambda_{\ell\, b}}, \label{Eqn:FirstBound}
\end{align}
where recall that the left-edge of the first partition $I^l_1 =
\lambda_{\ell \, b}$. Similarly we can show that $w.p.1$,
\begin{align}
\gamma_N &\geq \int_0^\infty \frac{1}{\lambda}d\Psi(\lambda) - \delta_1 - 2\frac{\delta_2}{\lambda_{\ell\, b}}\,. \label{Eqn:SecondBound}
\end{align}
Note that \eqref{Eqn:FirstBound} and \eqref{Eqn:SecondBound}
are similar in form to corresponding expressions in
\cite{TseHanly} except that the noise power $\sigma^2$ in
\cite{TseHanly} is replaced with $\lambda_{\ell \, b}$. The
remainder of the proof is identical to the proof of Lemma 4.3
in \cite{TseHanly} with $\sigma^2$ replaced by $\lambda_{\ell
\, b}$.
\subsection{Proof of Lemma \ref{Lemma:ErgodicSpatialAverage}}\label{Sec:ErgodicSpatialAverageLemmaProof}

To find $f_{P|\Pi_t}\left(P|\Pi_t\right)$, let  $\Xi_v$ be the
union of the set of disks of radius $v$ centered at each of the
base stations. Thus, for $x \leq P_M$, we have
\begin{align}
\Pr(P_i \leq x|\Pi_t) = \Pr(i\mbox{-th wireless-node } \in \Xi_v | \Pi_t)
\end{align}
with $v = \left(\frac{xG_t}{p_t}\right)^{1/\alpha}$. The set
$\Xi_v \backslash B(0,v)$ forms a Boolean model with radius $v$
disks as the primary grains (see Chapter 3 of \cite{Stoyan}).
The fractional area of the plane occupied by $\Xi_v\backslash
B(0,v)$ equals $1 - e^{-\rho_t \pi v^2}\, w.p.1.$
\cite{Stoyan}. Since $\Pr(i\mbox{-th wireless-node } \in B(0,v)
= 0)$, we have for $x \leq P_M$,
\begin{align}
\Pr(P_i \leq x|\Pi_t) = 1-e^{-\rho_t \pi  \left(\frac{xG_t}{p_t}\right)^{\frac{2}{\alpha}}}
\end{align}
and $\Pr(P_i \leq x|\Pi_t) = 1$ if $x > P_M$. Taking the
derivative with respect to $x$ yields
\eqref{Eqn:ErgodicPowerDens}. Taking the integral
$\int_0^\infty p^{\frac{2}{\alpha}} f_{P|\Pi_t}(p|\Pi_t)\,dp$ using
\eqref{Eqn:ErgodicPowerDens} yields
\eqref{Eqn:TetheredNodePowerScale}. When conditioned on $r_1$, the result still holds by the mixing property of the Poisson Voronoi tessallation (e.g. see \cite{okabe2009spatial}).

\subsection{Proof of Lemma \ref{Lemma:MinEval}} \label{Lemma:MinEvalProof}
Recall that $p_i = P_i G_t \,r_i^{-\alpha}$, and for $r_c > 0$
defined subsequently, write the matrix
\begin{align}
&\mathbf{K} = N^{{\frac{\alpha}{2}}-1} \sum_{i = 2}^{n+1} p_i \mathbf{g}_i\mathbf{g}_i^\dagger  \nonumber \\
&= \frac1N \sum_{i\in\mathcal{I}} N^{{\frac{\alpha}{2}}} P_i G_t  r_i^{-\alpha}  \mathbf{g}_i\mathbf{g}_i^\dagger + \frac1N \sum_{i\in\mathcal{I}^c}  N^{{\frac{\alpha}{2}}} P_i G_t  r_i^{-\alpha}  \mathbf{g}_i\mathbf{g}_i^\dagger\nonumber\\
&=\frac1N \sum_{i\in\mathcal{I}} N^{{\frac{\alpha}{2}}} \min\left(\frac{p_t r_c^\alpha}{G_t}, P_M\right) R^{-\alpha} G_t  \mathbf{g}_i\mathbf{g}_i^\dagger \nonumber \\
&+ \frac1N\sum_{i\in\mathcal{I}} N^{{\frac{\alpha}{2}}}
\left(P_ir_i^{-\alpha}-\min\left(\frac{p_t r_c^\alpha}{G_t},
P_M\right)R^{-\alpha}\right) G_t
\mathbf{g}_i\mathbf{g}_i^\dagger\nonumber\\
&+\frac1N
\sum_{i\in\mathcal{I}^c}  N^{{\frac{\alpha}{2}}} P_i G_t
r_i^{-\alpha}
\mathbf{g}_i\mathbf{g}_i^\dagger\label{Eqn:SplitNodes2}\\
&=\frac1N\sum_{i\in\mathcal{I}} \left(\frac{\pi \rho_t}{c}\right)^\alpha\min\left(\frac{p_t r_c^\alpha}{G_t}, P_M\right)  G_t \mathbf{g}_i\mathbf{g}_i^\dagger \nonumber \\
&+ \frac1N\sum_{i\in\mathcal{I}} N^{{\frac{\alpha}{2}}} \left(P_ir_i^{-\alpha}-\min\left(\frac{p_t r_c^\alpha}{G_t}, P_M\right)R^{-\alpha}\right) G_t    \mathbf{g}_i\mathbf{g}_i^\dagger \nonumber \\
&\;\;\;\;\;\;\;\;\;\;\;\;\;\;\;\;\;\;\;\;\;\;\;\;\;\;\;\;\;\;+\frac1N \sum_{i\in\mathcal{I}^c}  N^{{\frac{\alpha}{2}}} P_i G_t  r_i^{-\alpha}  \mathbf{g}_i\mathbf{g}_i^\dagger \label{Eqn:SplitNodes3}.
\end{align}
where $\mathcal{I} = \{i: r_{ti} < r_c, 1 < i \leq n+1\}$ is
the set of interferers that are closer than $r_c$ to their
closest base-stations. $\mathcal{I}^c = \{i: r_{ti} \geq
r_c, 1 < i \leq n+1\}$ is the set of remaining interferers. The
step from \eqref{Eqn:SplitNodes2} to \eqref{Eqn:SplitNodes3} is
from substituting $c = n/N$ and
\eqref{Eqn:NumberOfNodesRadius}.


For the hexagonal-cell model, with $A_H$ denoting the area of
each hexagonal cell, $r_c = \sqrt{ A_H/(2\pi)}\,$. 
For the Poisson-cell model, $r_c = \sqrt{{\ln
2}{(\rho_t\,\pi)}}$. These values are selected such that if we
place disks of radius $r_c$ around each base-station, the
fractional area occupied by the union of these disks is
$\frac12$, exactly for the hexagonal-cell model, and $w. p. 1$
for the Poisson-cell model (see Chapter 3 of \cite{Stoyan}).
Thus, as $n, N, R, \to\infty$, with probability approaching 1,
$\frac1N|\mathcal{I}|\to c/2$ and $\frac1N|\mathcal{I}^c|\to
c/2$.  Let the first term on the RHS of \eqref{Eqn:SplitNodes3}
be denoted by $\mathbf{T}_1\in\mathbb{C}^{N\times N}$. Note
from the system model that $r_i \leq R$ and
$P_ir_i^{-\alpha}-\min\left(\frac{p_t r_c^\alpha}{G_t},
P_M\right)R^{-\alpha} > 0$ for $i \in\mathcal{I}$ and $R$
sufficiently large. From \eqref{Eqn:SplitNodes3} and Weyl's
inequality (see e.g., \cite{Horn1990}),
$\lambda_{min}(\mathbf{K}) \geq \lambda_{min}(\mathbf{T}_1)$.
Additionally, from \cite{BaiSmallestEval}, as $n, N \to \infty$
such that $n/N \to c/2$,
\begin{align*}
 &\lambda_{min}(\mathbf{T}_1) \to \bar{\lambda}_{min}(\mathbf{T}_1) \nonumber \\
 &= \left(\frac{\pi \,\rho_t}{c}\right)^\alpha\,G_t\,\min\left(\frac{p_t\,r_c^\alpha}{G_t}, P_M\right)(1-\sqrt{2/c})^2 \;\; w. p. 1.
\end{align*} Hence, as $n,N, R\to\infty$, $w. p. 1$, the limiting e.d.f. of the eigenvalues of
$\mathbf{K}$ has support that is bounded from below by a
non-negative number $\lambda_{\ell b}$. Note that
$\lambda_{\ell b}$ could equal
$\frac12\bar{\lambda}_{min}(\mathbf{T}_1)$ for example.
Additionally, from \cite{BaiSilversteinNoMinimum}  for
sufficiently large $N$, $w. p. 1$, no eigenvalues of the matrix
$\mathbf{K}$ occur outside the support of the limiting e.d.f.
of the eigenvalues of $\mathbf{K}$. Hence, for $N$ sufficiently
large, there are no eigenvalues of $\mathbf{K}$ that are less
than $\lambda_{\ell\, b} > 0$  $w. p. 1$.

\subsection{Proof of Lemma \ref{Lemma:CellMild}} \label{Sec:CellMildProof}
Here, we show that $r_i/\sqrt{N}$ and $P_i$ are asymptotically independent for both hexagonal and Poisson cells. Recall that node $i$ is distributed with uniform probability in the radius $R$ circular network and
 $r_i$, $r_{ti}$ and $P_i$ are its distance to the origin, distance to its closest base station and transmit power respectively, 
and $\Xi_v$ is the union of the set of disks of radius $v$
centered at the base stations. Conditioned on $r_i \leq y
\sqrt{N}$, node-$i$ is uniformly distributed in
$B(0,{y\sqrt{N}})$. Hence,
\begin{align}
&\Pr\left\{ \left. r_{ti} \leq v \right| \frac{r_i}{\sqrt{N}} \leq y \right\} = \Pr\left\{ \left. r_{ti} \leq v \right| r_i \leq y \sqrt{N} \right\}\nonumber \\
& = \frac{\mbox{Area}\left(\Xi_v \cap B(0,{ y\sqrt{ N}})\right)}{\pi y^2 N} \label{Eqn:AreaFraction}
\end{align}
For hexagonal cells, as $N$ and $R\to\infty$, the RHS
approaches $F_X( v )$ from \eqref{Eqn:HexDistanceDist} as the
edge effects diminish. For Poisson cells, the set $\Xi_v
\backslash B(0,v)$ forms a Boolean model  and as $N \to\infty$,
the RHS of \eqref{Eqn:AreaFraction} converges to $1 -
e^{-\rho_t \pi v^2}$ $w. p. 1$ (see Chapter 3 of
\cite{Stoyan}).  Hence, $r_{ti}$ and $r_i/\sqrt{N}$ are
asymptotically independent and since $P_i$ is a function of
$r_{ti}$, $P_i$ is asymptotically independent of
$r_i/\sqrt{N}$. Thus, as $N\to\infty$, with probability
approaching unity,
\begin{align*}
&\Pr\left\{\left.\frac{r_{i}}{\sqrt{N}} \geq  \left(\frac{P_i}{x}\right)^{\frac{1}{\alpha}}\right|P_i \right\} \to \Pr\left\{\frac{r_{i}}{\sqrt{N}} \geq  \left(\frac{P_i}{x}\right)^{\frac{1}{\alpha}} \right\}\nonumber \\
&= \frac{R^2-N\left(\frac{P_i}{x}\right)^{\frac{2}{\alpha}}}{R^2}I_{ \left\{0<\left(\frac{P_i}{x}\right)^{-\frac{1}{\alpha}}\sqrt{N} < R\right\}}.
\end{align*}
Note that the RHS is simply the CDF of $r_i$ evaluated at $\sqrt{N}\left(\frac{P_i}{x}\right)^{\frac{1}{\alpha}}$.  Substituting $c = n/N$, \eqref{Eqn:NumberOfNodesRadius}, Bayes' rule, and rearranging terms in the RHS of the last equation yields \eqref{Eqn:PowerDistRest}.

\bibliography{IEEEabrv,main}

\end{document}